\documentclass[letterpaper]{article}

\usepackage{url}
\usepackage{graphicx}
\usepackage{soul}
\usepackage[utf8]{inputenc}
\usepackage[small]{caption}
\usepackage[svgnames]{xcolor}

\usepackage{multirow}

\usepackage{microtype}

\usepackage{tikz}
\usetikzlibrary{decorations.markings}
\usetikzlibrary{shapes.geometric,positioning,arrows}

\pgfdeclarelayer{edgelayer}
\pgfdeclarelayer{nodelayer}
\pgfsetlayers{edgelayer,nodelayer,main}
\tikzstyle{none}=[inner sep=0pt]
\tikzstyle{mystyle}=[circle,draw=black,scale=1.0,inner sep=0.5pt,font=\footnotesize,text width=4.75mm, align=center]

\newcommand{\shortcite}[1]{\cite{#1}}

\usepackage{pifont}
\usepackage{latexsym}
\usepackage{amsfonts}
\usepackage{amsmath}
\usepackage{amssymb}

\newcommand{\p}{{\rm P}}
\newcommand{\np}{{\rm NP}}
\newcommand{\conp}{{\rm coNP}}

\newcommand{\elec}{\ensuremath{\cal E}}
\newcommand{\thetatwo}{\ensuremath{\Theta_2^p}}

\newcommand{\deltatwo}{\ensuremath{{\rm P}^{\rm NP}}} %
\newcommand{\sigmatwo}{\ensuremath{\Sigma_2^{p}}}
\newcommand{\npnp}{\ensuremath{\np^\np}}

\newtheorem{theorem}{Theorem}
\newtheorem{corollary}[theorem]{Corollary}

\newtheorem{lemma}[theorem]{Lemma}
\newcommand\qedblob{\ding{113}}
\def\literalqed{{\ \nolinebreak\hfill\mbox{\qedblob\quad}}}

\newtheorem{observation}[theorem]{Observation}
\newtheorem{example}{Example}

\newenvironment{proofs}{\noindent{\bf Proof.}\hspace*{1em}}{\literalqed\bigskip}
\newenvironment{proofsketch}{\noindent{\bf Proof Sketch.}\hspace*{1em}}
{\literalqed\bigskip}

\showhyphens{}

\newcommand{\ccdcstar}{{{CCDC}$^*$}} %
\newcommand{\qsattwo}{\ensuremath{{\rm QSAT}_2}}

\newcommand{\prob}[3]{

\smallskip

\noindent
  {\bf Name:} #1
  
  \noindent
  {\bf Given:} #2
  
  \noindent
  {\bf Question:} #3
  
  \smallskip}

\begin{document}
\sloppy

\title{Very Hard Electoral Control Problems}

\author{Zack Fitzsimmons\\
Dept.\ of Math.\ and Comp.\ Science\\
College of the Holy Cross\\
Worcester, MA 01610
\and
Edith Hemaspaandra\\
Dept.\ of Computer Science\\
Rochester Institute of Technology\\
Rochester, NY 14623
\and
Alexander Hoover\\
Dept.\ of Computer Science\\
The University of Chicago\\
Chicago, IL 60637
\and
David E. Narv\'{a}ez\\
College of Comp.\ and Inf.\ Sciences\\
Rochester Institute of Technology\\
Rochester, NY 14623}

\date{November 13, 2018}

\maketitle

\begin{abstract}
It is important to understand how the outcome of an
election can be modified by an agent with control over the structure of the
election. Electoral control has been studied for many election systems, but
for all studied systems the winner problem is in P, and
so control is in NP.
There are election systems, such as Kemeny, that have many desirable properties,
but whose winner problems are not in NP.  Thus for such systems control is not in NP,
and in fact we show that it is typically complete for
\sigmatwo\ (i.e., $\npnp$, the second level of the polynomial
hierarchy).
This is a very high level of complexity. %
Approaches that perform quite well for solving NP problems do
not necessarily work for \sigmatwo-complete problems.
However, answer set programming is %
suited to express problems in
\sigmatwo, and we present an encoding for Kemeny control.
\end{abstract}

\section{Introduction}

The study of elections often deals with trade-offs for different properties
that an election system satisfies. Elections have a wide range of
applications from voting in political elections to applications in
artificial intelligence (see, e.g., Brandt et al.~\shortcite{bra-con-end-lan-pro:b:comsoc-handbook}).
And given the role of elections in multiagent system settings, it is
important that we study the computational properties of election systems.

Attacks on the structure of an election, referred to as control, were introduced by
Bartholdi, Tovey, and Trick~\shortcite{bar-tov-tri:j:control} and model natural scenarios
where an agent, with control over the structure of an election, %
modifies the structure (e.g., by adding candidates) to ensure that their preferred
candidate wins. It is important to study how 
these types of attacks
on the structure of an election can
 affect the outcome and how
computationally difficult it is to determine if such an attack exists. 

The complexity of electoral
control has been studied for many different
 natural elections systems %
and has been an important line of research in
computational social choice (see, e.g., Faliszewski and Rothe~\shortcite{fal-rot:b:handbook-comsoc-control-and-bribery}).
However,
for all of those systems the winner problems are in \p, and so the standard control
problems are in \np.
To go beyond what can easily be encoded for SAT solvers,
we need to look at election systems with harder winner problems.
But even for election systems whose winner problems are in \np, the most common control
problems are still in \np.

There are election systems that have many desirable social-choice properties, but whose
winner problems are not in \np\ (assuming $\np \neq \conp$). One important example
is the Kemeny rule~\cite{kem:j:no-numbers}, whose winner problem is
\thetatwo-complete (i.e., %
complete for parallel access to \np)~\cite{hem-spa-vog:j:kemeny} and so
the complexity of the standard control problems for Kemeny are all \thetatwo-hard
and thus also not in \np. For election systems with $\thetatwo$-complete winner problems, control is
in \sigmatwo\ (i.e., $\npnp$, the second level of the polynomial hierarchy), and we show
that control is typically \sigmatwo-complete for such systems.

This is a very high level of complexity. And so a natural question
to ask is how %
Kemeny control can be solved.
We mention here that there has been a long line of work that considers ways to
solve hard election problems. This includes work on
computing Kemeny winners %
(see, e.g.,~\cite{con-dav-kal:c:kemeny,bet-fel-guo-nie-ros:c:fpt-kemeny-aaim,ali-mei:j:experiments-kemeny,bet-bre-nie:j:theoretical-empirical-kemeny}) and on
solving election-attack problems %
(see, e.g.,~Rothe and Schend~\shortcite{rot-sch:j:typical-case-challenges}).
The work on solving hard election-attack problems has
been restricted to problems in \np, and such approaches do not work for \sigmatwo-complete problems.
Answer set programming (ASP) is an approach that has been recently applied for winner determination 
in voting, including for systems with hard winner problems~\cite{cha-pfa:c:democratix}.
ASP is suited to express \sigmatwo\ problems.
However, encoding \sigmatwo-complete problems in ASP requires the
use of more advanced techniques than encoding computationally easier problems.
We present an encoding of Kemeny control using these advanced techniques and
discuss other related approaches.

We make the following main contributions.
\begin{itemize}

\item We obtain the first natural \sigmatwo-completeness results for elections.

\item We define several new
natural and simple \sigmatwo-complete graph problems to prove our results,
and these
problems compare favorably
in naturalness and simplicity to other $\sigmatwo$-complete
problems, and so in usefulness as problems to reduce from.

\item We show for the most commonly-studied election systems with \thetatwo-complete winner problems, including the Kemeny rule, that control is typically
\sigmatwo-complete.

\item We build upon recent work on combining ASP with
   voting~\cite{cha-pfa:c:democratix} by encoding our \sigmatwo-complete
   control problems using advanced ASP techniques.
\end{itemize}

\section{Preliminaries}

An election consists of a set of candidates $C$, and a set of voters $V$, where
each voter has a vote that strictly ranks each candidate from most to least preferred.
An election system, \elec, maps an election $(C,V)$ to a set of winners, where the winner set can
be any subset of the candidate set. %
The winner problem for an election
system, \elec-Winner, is defined in the following way. Given an election $(C,V)$ and a candidate
$p \in C$, is $p$ a winner of the election using election system~\elec?

We consider the most important election systems with \thetatwo-complete winner problems:
Kemeny, Young, and Dodgson.

A candidate
is a Kemeny winner if it is the most-preferred candidate in a Kemeny
consensus~\cite{kem:j:no-numbers}, which
is a total order ``$>$'' that minimizes the sum of Kendall's Tau distances
(i.e., number of pairwise disagreements) with the voters in an election, i.e.,
minimizes $\sum_{a,b \in C, a > b} \|\{v \in V\ |\ b >_v a\}\|,$
where $>_v$ denotes the preference of voter $v$.

A candidate is a Young winner if it can become a weak Condorcet
winner (a candidate that beats-or-ties every
other candidate pairwise) by deleting the fewest voters~\cite{you:j:condorcet-theory}.

A candidate is a Dodgson winner if it can become a Condorcet winner
(a candidate that beats every other candidate pairwise) with the fewest swaps
between adjacent candidates in the voters' rankings~\cite{dod:unpubMAYBE-without-embedded-citations:dodgson-voting-system}.

Electoral control models the actions of an agent,
referred to as the election chair, who modifies the structure of the election (e.g.,
the voters) to ensure that their preferred candidate wins (in the constructive case)~\cite{bar-tov-tri:j:control} or that their despised candidate does
not win (in the destructive case)~\cite{hem-hem-rot:j:destructive-control}.\footnote{We
mention here that early work that considered electoral control generally
used the unique winner
model whereas we allow multiple winners. This rarely makes a difference
in the complexity for concrete systems.} (These two papers define the standard
control actions.)
We formally
define constructive control by adding candidates (CCAC) below, which models the real-world
scenario of an election chair adding spoiler candidates to an election to ensure that their preferred
candidate wins.

\prob{\elec-CCAC}{A set of registered candidates $C$, a set of unregistered candidates $D$,
a set of voters $V$ having preferences over $C \cup D$, an addition
limit $k$, and a preferred candidate $p \in C$.}{Does there
exist a set $D' \subseteq D$ such that $\|D'\| \le k$ and $p$ is a winner of
$(C \cup D', V)$ using election system~\elec?}

\noindent \paragraph{\bf Computational Complexity}
Our results concern the complexity classes $\thetatwo$ and $\sigmatwo$.
The class $\thetatwo$ was introduced in~\cite{pap-zac:c:two-remarks},
named in~\cite{wag:j:bounded}, and shown to be equivalent to 
$\p^\np_{||}$, 
the class of problems that can be solved by a polynomial-time machine with parallel access
to an \np\ oracle, in~\cite{hem:j:sky}.
$\sigmatwo = \npnp$ is the class of problems solvable by a nondeterministic polynomial-time machine with
access to an \np\ oracle, and is the second level of the
polynomial hierarchy~\cite{mey-sto:c:reg-exp-needs-exp-space,sto:j:poly}.
Note that $\np \cup \conp \subseteq \thetatwo \subseteq \p^\np \subseteq 
\sigmatwo$.

\section{Complexity Results}

 In this section we show that control problems for Kemeny, Young, and Dodgson elections are
typically \sigmatwo-complete.

\begin{observation}\label{obs:control}
For an election system \elec, the complexity of each standard
control action is in %
$\np^{\elec{\text{-}{\rm Winner}}}$.
\end{observation}

It is easy to see that the above observation holds. For a given election,
guess the control action of the chair (e.g., the set of candidates to be added)
and then use the oracle to check that the preferred candidate
is a winner (or that the despised candidate is not a winner).
In the case of partition control, which will not be discussed
further in this paper, the oracle will also be used to determine
which candidates participate in the runoff.
The winner problems for Kemeny, Young, and Dodgson 
are each in \thetatwo, and so the complexity of each standard control
action is in $\np^{\thetatwo} = \sigmatwo$.\footnote{Note that $\sigmatwo = \np^{\np} \subseteq \np^{\thetatwo} \subseteq \np^{\p^\np} = \np^\np = \sigmatwo$.}

\begin{corollary}
\label{cor:upperbound}
For Kemeny, Young, and Dodgson elections, the complexity of each standard
control action is in $\sigmatwo$.
\end{corollary}

As mentioned in~Brandt et al.~\shortcite{bra-bri-hem-hem:j:sp2}, these control problems
inherit \thetatwo-hardness from their winner problems. 

We will now show that these control problems are typically
$\sigmatwo$-complete.
Our $\sigmatwo$-completeness results are summarized
in Table~\ref{tbl:results} and we conjecture that
for Kemeny, Young, and Dodgson elections, the complexity of each standard
control action is $\sigmatwo$-complete.\footnote{De~Haan~\shortcite{haa:c:kemeny-judge-attacks} shows \sigmatwo-hardness
for control by adding/deleting issues for an analogue of Kemeny
for judgment aggregation.
Since our setting is much more restrictive,
this lower bound does not at all imply our lower bound.}

\begin{table}[h]
\centering
\caption{Summary of our \sigmatwo-completeness results for control.
Kemeny$'$ refers to a natural variant of
Kemeny defined in~\cite{dwo-kum-nao-siv:c:rank-aggregation} and
($*$) refers to the variant of control where the chair can delete
only certain candidates.}
\begin{tabular}{l c | c}
& Adding & Deleting \\
Voters & Young (Thm~\ref{thm:young:ccav}) &
\multirow{2}{*}{Young  (Thm~\ref{thm:young:ccdv})}\\
&  Kemeny$'$  (Thm~\ref{thm:kemprime:ccav}) & \\
\hline
Candidates & Kemeny  (Thm~\ref{thm:kem:ccac})
 & Kemeny ($*$) (Thm~\ref{thm:kem:ccdcstar})\\
& Dodgson (Thm~\ref{thm:dodgson:ccc})
& Dodgson (Thm~\ref{thm:dodgson:ccc})
\end{tabular}
\label{tbl:results}
\end{table}

We mention here that there are far fewer completeness results for
$\sigmatwo$ than there are for \np\ (see Schaefer and
Umans~\shortcite{sch-uma:j:PH-part-one,sch-uma:j:PH-part-two} for
a list of completeness results in the polynomial hierarchy).
An important reason why proving $\sigmatwo$-hardness is difficult
is the scarcity of ``simple'' $\sigmatwo$-complete problems to reduce from.
For example, scoring versions of Kemeny, Young, and Dodgson are 
proven NP-hard by reductions from simple NP-complete problems
such as Vertex-Cover, but prior to this paper there were no $\sigmatwo$-complete simple
versions of Vertex-Cover that were suitable to show that related
control-by-adding problems are $\sigmatwo$-hard.

Below we define simple and natural $\sigmatwo$-complete versions of 
Vertex-Cover (and the analogous Independent-Set versions
are also $\sigmatwo$-complete).
We will see that these problems are particularly useful to show
that our control problems are $\sigmatwo$-hard.
Of course, we need to show that our new simple problems are
$\sigmatwo$-hard, which is difficult. But we can then reuse
our simple problems to obtain simpler $\sigmatwo$-hard
proofs for multiple other problems.

The following problem (and its closely related Independent-Set
analogue) is particularly useful to reduce to control-by-adding
problems. For example, we will see that this problem quite easily
reduces to Kemeny-CCAC and that the Independent-Set analogue of this
problem reduces quite easily to Young-CCAV (control by
adding voters). 

\prob{Vertex-Cover-Member-Add}
{Graph $G = (V \cup V',E)$, set of addable vertices $V'$,
addition limit $k$, and vertex $\hat{v} \in V$.}
{Does there exist a set $W \subseteq V'$ of at most $k$ addable vertices
such that
$\hat{v}$ is a member of a minimum vertex cover\footnote{A vertex cover
of a graph is a set of vertices such that every edge is incident with at
least one vertex in the set.}
of $(V \cup W,E)$?}

\begin{theorem}\label{thm:vcma}
Vertex-Cover-Member-Add is $\sigmatwo$-complete.
\end{theorem}

To show the essence of the argument of the proof of
Theorem~\ref{thm:vcma} and avoid some of the more finicky details
of the proof, we prove that
Vertex-Cover-Member-Select is \sigmatwo-complete, and then
briefly discuss how this proof can be adapted for
Vertex-Cover-Member-Add.\footnote{\label{f:GND}We can easily
prove Vertex-Cover-Member-Select $\sigmatwo$-hard by reducing
the $\sigmatwo$-complete problem
Generalized-Node-Deletion to it, in which
we are given a graph, and two
integers $k$ and $\ell$, and we ask if we can delete at most $k$ vertices such
that the remaining graph does not contain $K_{\ell + 1}$, a clique of size
$\ell+1$~\cite{rut:j:prop-truth-maintenance}.
For the reduction, simply output $(H,V(H),k,\hat{v})$, where $H$ is the graph
$(\overline{G} \cup (\{\hat{v}\},\emptyset)) \times \overline{K_{\ell+1}}$.

However, this proof does not generalize to 
Vertex-Cover-Member-Add. In particular, the ``adding'' analogue
of Generalized-Node-Deletion in which we ask if we can \emph{add}
up to $k$ vertices such that the resulting graph does not
have a clique of size $\ell + 1$, is in P (since it is best to add
nothing). And the version where we ask if there is a clique of size
$\ell + 1$ after adding is in NP.}

\prob{Vertex-Cover-Member-Select}
{Graph $G = (V,E)$, a set $V' \subseteq V$ of deletable vertices,
delete limit $k$, and vertex $\hat{v} \in V$.}
{Does there exist a set $W \subseteq V'$ of at most $k$ deletable vertices
such that
$\hat{v}$ is a member of a minimum vertex cover of $G - W$?\footnote{$G - W = (V(G) - W, \{\{v, w\}~|~\{v, w\} \in E(G) \wedge v,w \in V(G)-W\})$.}}

\begin{lemma}\label{l:vcms}
Vertex-Cover-Member-Select is $\sigmatwo$-complete.
\end{lemma}

\begin{proofs}
Membership in $\sigmatwo$ is easy to see: Guess at most $k$ deletable 
vertices to delete, then guess a vertex cover containing $\hat{v}$ and use
the \np\ oracle to check that the guessed vertex cover is a minimum vertex 
cover.

To show hardness, we reduce from the following
$\sigmatwo$-complete problem, QSAT$_2$~\cite{sto:j:poly,wra:j:complete}:
all true formulas of the form $\exists x_1 \cdots \exists x_n
\neg(\exists y_1 \cdots \exists y_n 
\phi(x_1, \ldots, x_n, y_1, \ldots, y_n))$,
where
$\phi$ is a formula in 3cnf.\footnote{Note that we have the same number of
variables in each quantified block (we can simply pad to get this). Also,
we pull the negation out of the formula so that the formula is in 3cnf
and not in 3dnf.}

We first recall the standard reduction from 3SAT to
Vertex-Cover~\cite{kar:b:reducibilities}.
Let $G$ be the graph constructed by this  
reduction
on $\phi(x_1, \ldots, x_n, y_1, \ldots, y_n)$, where $\phi$ is in 3cnf.
Let
$\phi = \psi_1 \wedge \psi_2 \wedge \cdots \wedge \psi_m$ and for each $i, 1 \le i \le m$,
$\psi_i = c_{i,1} \vee c_{i,2} \vee c_{i,3}$.
Graph $G$ consists of $4n+3m$ vertices: a vertex for each $x_i$, $\overline{x_i}$,
$y_i$, and $\overline{y_i}$ and for each clause $i, 1 \le i \le m$, three vertices $c_{i,1}$, $c_{i,2}$, and $c_{i,3}$,
and the following edges:
\begin{itemize}
\item
for each $i, 1 \le i \le n$, the edges $\{x_i,\overline{x_i}\}$ and
$\{y_i, \overline{y_i}\}$,
\item
for each $i, 1 \le i \le m$, the edges $\{c_{i,1}, c_{i,2}\}, \{c_{i,1}, c_{i,3}\}$,
and $\{c_{i,2},c_{i,3}\}$, (i.e., the complete graph on three vertices),
\item and for each 
vertex $c_{i,j}$ we have an edge to its corresponding vertex candidate (e.g.,
if $c_{i,j} = x_t$ in $\phi$ then we have the edge $\{c_{i,j}, x_t\}$).
\end{itemize}
An example of this construction will follow.

Note that every vertex cover of $G$ contains at least one
of each $\{x_i,\overline{x_i}\}$, at least one of each $\{y_i,\overline{y_i}\}$,
and at least two of each $\{c_{i,1},c_{i,2},c_{i,3}\}$,
so $G$ does not have a vertex cover of size less than $2n+2m$.
The properties below follow immediately from the proof
from~\cite{kar:b:reducibilities}.
\begin{enumerate}
\item If $X$ is a vertex cover of size $2n + 2m$, then
$X \cap \{x_i, \overline{x_i}, y_i, \overline{y_i}  \ | \ 1 \leq i \leq n\}$
corresponds to a satisfying assignment for $\phi$.
\item If $\alpha$ is a satisfying assignment for $\phi$, then
there is a vertex cover $X$ of size $2n+2m$ such that
$X \cap \{x_i, \overline{x_i}, y_i, \overline{y_i} \ | \ 1 \leq i \leq n\}$
corresponds to this assignment.
\end{enumerate}

Below we include an example of this construction given the formula
$\phi = (x_1 \vee x_1 \vee y_1) \wedge (\overline{x_1} \vee \overline{y_1} \vee \overline{y_1})
\wedge (\overline{x_1} \vee y_1 \vee y_1)$.
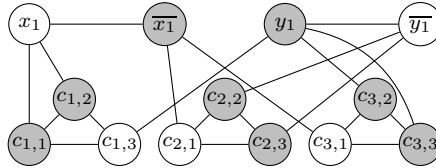
{

\centering
\begin{tikzpicture}[scale=0.8]
	\begin{pgfonlayer}{nodelayer}
		\node [style=mystyle,fill=lightgray] (0) at (-2.5, 3) {$\overline{x_1}$};
		\node [style=mystyle,fill=lightgray] (1) at (-0.5, 3) {$y_1$};
		\node [style=mystyle] (2) at (1.75, 3) {$\overline{y_1}$};
		
		\node [style=mystyle,fill=lightgray] (3) at (-4, 1.75) {$c_{1,2}$}; %
		\node [style=mystyle,fill=lightgray] (4) at (-4.75, 1) {$c_{1,1}$}; %
		\node [style=mystyle] (5) at (-3.25, 1) {$c_{1,3}$}; %
		
		\node [style=mystyle,fill=lightgray] (6) at (-1.5, 1.75) {$c_{2,2}$}; %
		\node [style=mystyle] (7) at (-2.25, 1) {$c_{2,1}$}; %
		\node [style=mystyle,fill=lightgray] (8) at (-0.75, 1) {$c_{2,3}$}; %
		
		\node [style=mystyle,fill=lightgray] (9) at (1, 1.75) {$c_{3,2}$}; %
		\node [style=mystyle,fill=lightgray] (10) at (1.75, 1) {$c_{3,3}$}; %
		\node [style=mystyle] (11) at (0.25, 1) {$c_{3,1}$}; %
		
		\node [style=mystyle] (12) at (-4.75, 3) {$x_1$};
	\end{pgfonlayer}
	\begin{pgfonlayer}{edgelayer}
		\draw (1) to (2);
		\draw (4) to (3);
		\draw (3) to (5);
		\draw (6) to (7);
		\draw (8) to (6);
		\draw (9) to (11);
		\draw (10) to (11);
		\draw (8) to (7);
		\draw (5) to (4);
		\draw [in=180, out=0, looseness=1.00] (12) to (0);
		\draw (12) to (3);
		\draw (12) to (4);
		\draw (1) to (5);
		\draw (0) to (7);
		\draw (0) to (11);
		\draw (2) to (6);
		\draw (2) to (8);
		\draw [bend left, looseness=1.00] (1) to (10);
		\draw (9) to (10);
		\draw (1) to (9);
	\end{pgfonlayer}
\end{tikzpicture}

}

In the figure above vertices that are in a minimum vertex cover are shaded in gray, and
this corresponds to the satisfying assignment $x_1 = 0$, $y_1 = 1$
for $\phi$.

For the reduction from QSAT$_2$ to 
Vertex-Cover-Member-Select, we construct the graph $H$, which 
is a modified version of the graph $G$. For each clause $i, 1 \le i \le m$, instead of the
complete graph on three vertices, $\{c_{i,1},c_{i,2},c_{i,3}\}$, we add an extra vertex $d_i$ and
have the complete graph on four vertices, $\{c_{i,1},c_{i,2},c_{i,3},d_i\}$, and
we connect the fourth vertex $d_i$ of each clause gadget to a special new
vertex $\hat{v}$. So our graph $H$ consists of $4n+4m+1$ vertices and the
edges
as just described.
Below we give the graph corresponding to the same formula as the previous example.

Note that every vertex cover of $H$ contains at least one
of each $\{x_i,\overline{x_i}\}$, %
at least one of each $\{y_i,\overline{y_i}\}$, %
and at least three of each $\{c_{i,1},c_{i,2},c_{i,3},d_i\}$. %
So $H$ does not have a vertex
cover of size less than $2n+3m$, and there is a vertex cover of size 
$2n + 3m + 1$ that includes~$\hat{v}$.
Note that $H$ has the following properties.

\begin{enumerate}
\item If $X$ is a vertex cover of size $2n + 3m$, then
 $\hat{v} \not \in X$ and
 $X \cap \{x_i, \overline{x_i}, y_i, \overline{y_i}  \ | \ 1 \leq i \leq n\}$
 corresponds to a satisfying assignment for $\phi$.
 \item If $\alpha$ is a satisfying assignment for $\phi$, then
 there is a vertex cover of size $2n+3m$ such that
 $X \cap \{x_i, \overline{x_i}, y_i, \overline{y_i} \ | \ 1 \leq i \leq n\}$
 corresponds to this assignment.
\end{enumerate}

Below we include an example of this construction given the formula
$\exists x_1 \neg (\exists y_1 \phi(x_1,y_1))$, where as before\\ $\phi = (x_1 \vee x_1 \vee y_1) \wedge
(\overline{x_1} \vee \overline{y_1} \vee \overline{y_1}) \wedge (\overline{x_1} \vee y_1 \vee y_1)$.
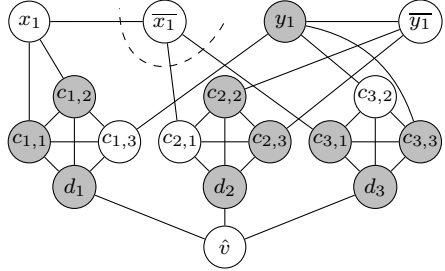
{

\centering
\begin{tikzpicture}[scale=0.8]
	\begin{pgfonlayer}{nodelayer}
		\node [style=mystyle] (0) at (-2.5, 3) {$\overline{x_1}$};
		\node [style=mystyle,fill=lightgray] (1) at (-0.5, 3) {$y_1$};
		\node [style=mystyle] (2) at (1.75, 3) {$\overline{y_1}$};
		
		\node [style=mystyle, fill=lightgray] (3) at (-4, 1.75) {$c_{1,2}$}; %
		\node [style=mystyle, fill=lightgray] (4) at (-4.75, 1) {$c_{1,1}$}; %
		\node [style=mystyle,fill=lightgray] (5) at (-4, 0.25) {$d_1$};
		\node [style=mystyle] (6) at (-3.25, 1) {$c_{1,3}$}; %
		
		\node [style=mystyle, fill=lightgray] (7) at (-1.5, 1.75) {$c_{2,2}$}; %
		\node [style=mystyle] (8) at (-2.25, 1) {$c_{2,1}$}; %
		\node [style=mystyle,fill=lightgray] (9) at (-1.5, 0.25) {$d_2$};
		\node [style=mystyle, fill=lightgray] (10) at (-0.75, 1) {$c_{2,3}$}; %
		
		\node [style=mystyle] (11) at (1, 1.75) {$c_{3,2}$}; %
		\node [style=mystyle, fill=lightgray] (12) at (1.75, 1) {$c_{3,3}$}; %
		\node [style=mystyle,fill=lightgray] (13) at (1, 0.25) {$d_3$};
		\node [style=mystyle, fill=lightgray] (14) at (0.25, 1) {$c_{3,1}$}; %
		\node [style=mystyle] (15) at (-1.5, -0.75) {$\hat{v}$};
		
		\node [style=mystyle] (16) at (-4.75, 3) {$x_1$};
		\node [style=none] (17) at (-1.5, 3) {};
		\node [style=none] (18) at (-3.25, 3.25) {};
	\end{pgfonlayer}
	\begin{pgfonlayer}{edgelayer}
		\draw (1) to (2);
		\draw (4) to (3);
		\draw (3) to (6);
		\draw (6) to (5);
		\draw (5) to (4);
		\draw (7) to (8);
		\draw (8) to (9);
		\draw (9) to (10);
		\draw (10) to (7);
		\draw (11) to (14);
		\draw (14) to (13);
		\draw [in=-135, out=45, looseness=1.00] (13) to (12);
		\draw (11) to (13);
		\draw (12) to (14);
		\draw (10) to (8);
		\draw (7) to (9);
		\draw (3) to (5);
		\draw (6) to (4);
		\draw (5) to (15);
		\draw (9) to (15);
		\draw (13) to (15);
		\draw [in=180, out=0, looseness=1.00] (16) to (0);
		\draw (16) to (3);
		\draw (16) to (4);
		\draw (1) to (6);
		\draw (0) to (8);
		\draw (0) to (14);
		\draw (2) to (7);
		\draw (2) to (10);
		\draw [bend left, looseness=1.00] (1) to (12);
		\draw (11) to (12);
		\draw (1) to (11);
		\draw [dashed, in=-90, out=-120, looseness=1.75] (17) to (18);
	\end{pgfonlayer}
\end{tikzpicture}

}

Note that in the figure when $\overline{x_1}$ is removed (i.e., setting $x_1 = 0$) a
minimum-size vertex cover (shaded in gray) is of size $n + 3m = 10$ and that $\phi(0,y_1)$
is satisfied with $y_1 = 1$.

We have repeated the same graph below, except now the vertex $x_1$ is removed
(i.e., setting $x_1 = 1$).
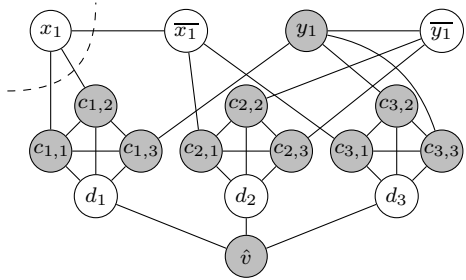
{

\centering
\begin{tikzpicture}[scale=0.8]
	\begin{pgfonlayer}{nodelayer}
		\node [style=mystyle] (0) at (-2.5, 3) {$\overline{x_1}$};
		\node [style=mystyle,fill=lightgray] (1) at (-0.5, 3) {$y_1$};
		\node [style=mystyle] (2) at (1.75, 3) {$\overline{y_1}$};
		
		\node [style=mystyle,fill=lightgray] (3) at (-4, 1.75) {$c_{1,2}$}; %
		\node [style=mystyle,fill=lightgray] (4) at (-4.75, 1) {$c_{1,1}$}; %
		\node [style=mystyle] (5) at (-4, 0.25) {$d_1$};
		\node [style=mystyle,fill=lightgray] (6) at (-3.25, 1) {$c_{1,3}$}; %
		
		\node [style=mystyle,fill=lightgray] (7) at (-1.5, 1.75) {$c_{2,2}$}; %
		\node [style=mystyle,fill=lightgray] (8) at (-2.25, 1) {$c_{2,1}$}; %
		\node [style=mystyle] (9) at (-1.5, 0.25) {$d_2$};
		\node [style=mystyle,fill=lightgray] (10) at (-0.75, 1) {$c_{2,3}$}; %
		
		\node [style=mystyle,fill=lightgray] (11) at (1, 1.75) {$c_{3,2}$}; %
		\node [style=mystyle,fill=lightgray] (12) at (1.75, 1) {$c_{3,3}$}; %
		\node [style=mystyle] (13) at (1, 0.25) {$d_3$};
		\node [style=mystyle,fill=lightgray] (14) at (0.25, 1) {$c_{3,1}$}; %
		
		\node [style=mystyle,fill=lightgray] (15) at (-1.5, -0.75) {$\hat{v}$};
		\node [style=mystyle] (16) at (-4.75, 3) {$x_1$};
		\node [style=none] (17) at (-5.5, 2) {};
		\node [style=none] (18) at (-4, 3.5) {};
	\end{pgfonlayer}
	\begin{pgfonlayer}{edgelayer}
		\draw (1) to (2);
		\draw (4) to (3);
		\draw (3) to (6);
		\draw (6) to (5);
		\draw (5) to (4);
		\draw (7) to (8);
		\draw (8) to (9);
		\draw (9) to (10);
		\draw (10) to (7);
		\draw (11) to (14);
		\draw (14) to (13);
		\draw [in=-135, out=45, looseness=1.00] (13) to (12);
		\draw (11) to (13);
		\draw (12) to (14);
		\draw (10) to (8);
		\draw (7) to (9);
		\draw (3) to (5);
		\draw (6) to (4);
		\draw (5) to (15);
		\draw (9) to (15);
		\draw (13) to (15);
		\draw [in=180, out=0, looseness=1.00] (16) to (0);
		\draw (16) to (3);
		\draw (16) to (4);
		\draw (1) to (6);
		\draw (0) to (8);
		\draw (0) to (14);
		\draw (2) to (7);
		\draw (2) to (10);
		\draw [bend left, looseness=1.00] (1) to (12);
		\draw (11) to (12);
		\draw (1) to (11);
		\draw [dashed,bend right=45, looseness=1.25] (17) to (18);
	\end{pgfonlayer}
\end{tikzpicture}

}
The vertices shaded in gray above correspond to a minimum-size vertex
cover of size $n+3m+1 = 11$. Note that $\phi(1,y_1)$ is not satisfiable
and that this vertex cover includes $\hat{v}$.

We will show that $\exists x_1 \cdots \exists x_n
\neg (\exists y_1 \cdots \exists y_n \phi(x_1, \ldots,\linebreak[1] x_n,\linebreak[1]
y_1, \ldots, y_n))$
if and only if we can delete at most $n$ vertices in 
$\{x_i, \overline{x_i} \ | \ 1 \leq i \leq n\}$ such that
$\hat{v}$ is a member of a minimum vertex cover of $H$-after-deletion.

From the listed properties of $H$ and the above example, it is not too
hard to see
that the statement above holds as long as the vertices deleted from $H$ correspond
to an assignment to the $x$-variables. However, it is possible for the set of deleted vertices
to contain neither or both of $\{x_i,\overline{x_i}\}$. We handle these cases below.

Suppose that $W$ is a set of at most $n$ vertices from
$\{x_i, \overline{x_i} \ | \ 1 \leq i \leq n\}$ such that
$\hat{v}$ is a member of a minimum vertex cover of $H - W$.
Let $\widehat{n} \leq n$ be the number of 
$\{x_i, \overline{x_i}\}$ pairs that are undeleted, i.e.,
for which $\{x_i, \overline{x_i}\} \cap W = \emptyset$.
Note that the size of a vertex cover of $H - W$ is 
at least $\widehat{n} + n + 3m$, and that vertex covers of that
size do not include $\hat{v}$. Since $\hat{v}$ is a member of a minimum size
vertex cover of $H-W$, it follows that
$H - W$ does not have a vertex cover of size
$\widehat{n} + n + 3m$.

Let $\alpha \in \{0,1\}^n$ be an assignment to $x_1 \cdots x_n$ that is
consistent with $W$, in the sense that
for all $i, 1 \leq i \leq n$,
if $W \cap \{x_i,\overline{x_i}\} = \{x_i\}$, then $\alpha_i = 1$ and
if $W \cap \{x_i,\overline{x_i}\} = \{\overline{x_i}\}$, then $\alpha_i = 0$.
To complete the proof, suppose for a contradiction that
$\phi(\alpha_1, \ldots, \alpha_n, y_1, \ldots, y_n)$ is 
satisfiable. Then $H$ has a vertex cover $X$ of size
$2n+3m$ such that $X \cap 
\{x_i, \overline{x_i} \ | \ 1 \leq i \leq n\}$ corresponds
to $\alpha$. Note that $X - W$ is a vertex cover of $H - W$.
But $\|X \cap W\| = n - \widehat{n}$, and so 
$\|X - W\| = 2n + 3m - (n - \widehat{n}) =
\widehat{n} + n + 3m$,
which is a contradiction.~\end{proofs}

To prove Theorem~\ref{thm:vcma},
we use a similar reduction %
to show that Vertex-Cover-Member-Add is
$\sigmatwo$-hard.
The main difference is that we need an
edge and two vertices for each $x_i$ and for each $\overline{x_i}$.
For the proof, see the appendix. %

\medskip

Vertex-Cover-Member-Select reduces to the corresponding Kemeny control problem
Kemeny-\ccdcstar.

\prob{\elec-\ccdcstar}{An election $(C,V)$, a set of deletable candidates
$D \subseteq C$, a delete limit $k$, and a preferred candidate
$p \in C$.}{Does there exist a set $D' \subseteq D$ of at most $k$ deletable
candidates such that $p$ is a winner of $(C-D',V)$ using election system $\elec$?}

Note that \elec-\ccdcstar\ is more structured than \elec-CCDC (where the set of deletable
candidates is $C$),
since the chair can only delete from a subset of $C$. %

\begin{theorem}\label{thm:kem:ccdcstar}
Kemeny-\ccdcstar is $\sigmatwo$-complete.\footnote{%
This result holds even for four voters, 
using the construction from
Dwork et al.~\shortcite{dwo-kum-nao-siv:c:rank-aggregation}.
For details, see the appendix.} %
\end{theorem}

\begin{proofsketch}
Kemeny-Score
was shown to be \np-hard by a reduction from
Feedback-Arc-Set\footnote{A feedback arc set of a directed
graph is a set of arcs such that deleting this set makes
the graph acyclic.} by Bartholdi, Tovey, and Trick~\shortcite{bar-tov-tri:j:who-won},
which was shown to be \np-hard by Karp~\shortcite{kar:b:reducibilities}
by a reduction from Vertex-Cover. Both these reductions
are straightforward. To show the $\thetatwo$-hardness of
Kemeny-Winner, Hemaspaandra, Spakowski, and
Vogel~\shortcite{hem-spa-vog:j:kemeny} define $\thetatwo$-complete
versions of Vertex-Cover and Feedback-Arc-Set, namely Vertex-Cover-Member
and Feedback-Arc-Set-Member.
They show that Vertex-Cover-Member is \thetatwo-complete.
They then show that Vertex-Cover-Member reduces to
Feedback-Arc-Set-Member, which then reduces to Kemeny-Winner.
These two reductions are similar to the NP reductions and also
straightforward.
The same happens in our $\sigmatwo$ case:
Vertex-Cover-Member-Select easily and straightforwardly
reduces to Feedback-Arc-Set-Member-Select, which easily and
straightforwardly reduces to Kemeny-\ccdcstar.
For details, see the appendix. %
\end{proofsketch}

Vertex-Cover-Member-Add easily and similarly reduces
to Feedback-Arc-Set-Member-Add
(defined in the appendix) and then %
to Kemeny-CCAC.

\begin{theorem}
\label{thm:kem:ccac}
Kemeny-CCAC is $\sigmatwo$-complete.
\end{theorem}

When we try to similarly show that Kemeny-CCAV is $\sigmatwo$-complete, it
is easy to show that Feedback-Arc-Set-Member-Add-Arcs, where we add
arcs instead of vertices, is $\sigmatwo$-complete.
The problem is that in the reduction from Feedback-Arc-Set to Kemeny-Score,
each arc corresponds to two voters.  However, we were able to
show this result
for what we here call Kemeny$'$,
the natural variant of Kemeny from~\cite{dwo-kum-nao-siv:c:rank-aggregation} 
where the voters do not necessarily
list all of the candidates in their votes and
unlisted candidates in a vote do not contribute
to the distance to the Kemeny consensus and so do not increase the
Kemeny score. In this case, one arc will correspond to one voter.

\begin{theorem}
\label{thm:kemprime:ccav}
Kemeny$'$-CCAV is $\sigmatwo$-complete.
\end{theorem}

We now turn to Young elections. We will explain how %
Independent-Set-Member-Delete,
the Independent-Set analogue of Vertex-Cover-Member-Delete,
which is also $\sigmatwo$-complete,
is useful to show %
Young-CCDV is $\sigmatwo$-complete.

\prob{Independent-Set-Member-Delete}
{Graph $G = (V,E)$,
delete limit $k$, and vertex $\hat{v} \in V$.}
{Does there exist a set $W \subseteq V$ such that $\|W\| \le k$ and
$\hat{v}$ is a member of a maximum independent set of $G - W$?}

\begin{theorem}\label{thm:ismd}
Independent-Set-Member-Delete is $\sigmatwo$-complete.
\end{theorem}
The proof of the above theorem can be
found in the appendix.
We reduce this
problem to Young-CCDV to get the following result.

\def\ismdproof{
\begin{proofs}
Independent-Set-Member-Delete is clearly in \sigmatwo.
We will show that Independent-Set-Member-Delete is 
$\sigmatwo$-complete by reducing the $\sigmatwo$-complete problem
Generalized-Node-Deletion to it, in which we are given a graph $G$, and two
integers $k$ and $\ell$, and we ask if we can delete at most $k$ vertices such
that the remaining graph does not contain $K_{\ell + 1}$, a clique of size
$\ell+1$~\cite{rut:j:prop-truth-maintenance}.
For the reduction, simply output $(H,k,\hat{v})$, where $H$ is the graph
$\overline{G}  \times \overline{K_{\ell}}$, and $\hat{v}$ is a vertex in
$\overline{K_{\ell}}$.

If we can delete at most $k$ vertices in $G$ such that the remaining 
graph does not contain a clique of size $\ell + 1$, then
$\overline{G}$ after deletion does not contain an independent set
of size $\ell + 1$. It follows that after deletion,
$\overline{K_{\ell}}$ is a largest 
independent set in $H$ and $\hat{v}$ is an element of this independent set.

For the converse, suppose we delete at most $k$ vertices from $H$ such that
$\hat{v}$ is a member of a largest independent set. This independent
set has size at most $\ell$, and so after
deletion, $\overline{G}$ does not have an independent set of size
$\ell + 1$.
\end{proofs}}

\begin{theorem}
\label{thm:young:ccdv}
Young-CCDV is \sigmatwo-complete.
\end{theorem}

\def\youngccdv{
\begin{proofs}
For $G$ a graph, $\alpha(G)$
is the independence number of $G$, i.e., the size of a
maximum independent set of $G$. For $v$ a vertex,
$\alpha_v(G)$ is the size of a maximum independent set of 
$G$ that contains $v$.

Given a graph $G = (V,E)$, a delete limit $k \leq \|V\|$, and a vertex
$\hat{v} \in V$, below we construct an election with two special candidates
$p$ and $q$ such that our instance is a positive instance
of Independent-Set-Member-Delete if and
only if we can make the Young score\footnote{The Young score of a candidate
is the size of a largest subset of the voters that make it a weak
Condorcet winner. A Young winner is a candidate with highest Young score.}
of $p$ at least 2 and
at least as high as the Young score of $q$ by deleting at most
$k$ voters.

This is not quite the same as making $p$ a Young winner,
since other candidates could have higher scores and since it
is also possible for a Young winner to have a score below 2.
However, this can easily be handled as follows:
We can use the trick from~\cite{rot-spa-vog:j:young} to change the election
such that the Young scores of $p$ and $q$ remain the same,
the Young scores of all other candidates are at most 2, and
such that there is a candidate with a Young score of at least 2.

We use candidate $q$ to witness the size of a maximum independent set.
The main idea on how to do this is implicit in the construction from
Rothe, Spakowski, and Vogel~\shortcite{rot-spa-vog:j:young}.
The new twist is to use candidate $p$ 
to witness the size of a maximum independent set containing
$\hat{v}$.  In order to do so, we make sure that all 
voters corresponding to vertices connected to $\hat{v}$ are not in a
set of voters that realizes the Young score of $p$,
by making these voters maximally unattractive to $p$, by ranking
$p$ last.

Given a graph $G = (V,E)$, a delete limit $k \leq \|V\|$, and a vertex
$\hat{v} \in V$, without loss of generality
assume that $G$ has no isolated vertices
and that for every set $W$ of at most
$k$ vertices, $\alpha(G-W) \geq 3$ (the latter property can 
for example be ensured by adding two ($k+1$)-cliques to $G$).
Let the candidate set be $E \cup \{a,p,q\}$ and let the voter set consist of:

\begin{description}
\item[Type~IA] For each $v \in V$ such that $\{v,\hat{v}\} \not \in E$,
one voter corresponding to $v$ voting\\
$(\{e \in E \ | \ v \in e \} > a > q  > p > \cdots)$.
\item[Type~IB] For each $v \in V$ such that $\{v,\hat{v}\} \in E$,
one voter corresponding to $v$ voting\\
$(\{e \in E \ | \ v \in e \} > a > q  > \cdots > p)$.
\item[Type~II] 
One voter voting $(p > q > \cdots > a)$.
\item[Type~III] 
$2\|V\|$ voters voting $(\cdots > p > q > a)$.
\end{description}

Suppose $X$ is a set of at most $k$ voters that we delete.
If $X$ contains the voter of Type II, the Young scores
of $p$ and $q$ are 0 (since there are no isolated vertices).
If not, let $W$ be the set of of vertices corresponding
to the Type-I voters in $X$.
Careful and tedious inspection shows that the Young score of $q$ is
$2 \alpha(G - W)$ (this Young score is realized by a set of voters whose Type-I
voters correspond to a maximum independent set) and that the
Young score of $p$ is $2 \alpha_{\hat v}(G - W)$ 
(this Young score is realized by a set of voters whose Type-I
voters are all of Type-IA and correspond to a maximum independent set
that contains $\hat{v}$). For details, see the section below. %
\end{proofs}}

The proof of the above theorem can be found in the appendix. %
The same construction as used in this proof
gives a reduction %
to Young-CCAV.
\begin{theorem}
\label{thm:young:ccav}
Young-CCAV is \sigmatwo-complete.
\end{theorem}

Previous complexity results for Dodgson elections do not reduce from
problems related to Vertex-Cover or Independent-Set. 
However, in Dodgson there is more flexibility
in how to construct the
voters in a reduction, and so we were able to directly reduce
QSAT$_2$ to Dodgson-CCDC and Dodgson-CCAC, though the constructions
are quite involved.
The proofs can be found in the appendix. %
\begin{theorem}
\label{thm:dodgson:ccc}
Dodgson-CCDC and CCAC are \sigmatwo-complete.
\end{theorem}

\section{Encoding Control Problems with ASP}
\label{sec:asp}

When faced with a computationally difficult problem at the \np\ level there
are several different possible ways to encode the problem to harness the power
of solvers for hard problems. For example, problems in \np\ are often encoded
for Boolean satisfiability solvers. When a problem is \sigmatwo-complete there
are far fewer tools to use. However, we can encode our program for
an answer set programming (ASP) solver.

Answer set programming is a paradigm for encoding computationally 
difficult problems in a declarative way~(see, e.g., Brewka et al.~\shortcite{bre-eit-tru:c:asp-cacm}).
Using modern ASP input languages like the one 
in the Gringo grounder~\cite{gebser:15}, which extends conventional ASP 
with aggregates functions like {\tt \#sum} and {\tt \#count},
we can use variable 
names and predicates. A descriptive naming scheme usually leads to natural 
encodings of problems when
compared to other approaches such as
encoding into 
Boolean satisfiability problems.

Using ASP to solve computational problems in voting was first
proposed by Konczak~\shortcite{kon:c:asp-possible-necessary}.
Recent work by Charwat and Pfandler~\shortcite{cha-pfa:c:democratix} provides
winner-problem encodings for many election systems, including systems with hard 
winner problems, and mentions encoding control problems as future work.
The predicates used in their encodings are arguably 
self-explanatory and the use of aggregates provides succinct representations of 
the different
voting rules they consider.

Since encoding a problem in ASP can lead to natural encodings of a problem
and since ASP can encode problems in~\sigmatwo~\cite{eiter:97}, we discuss how
to encode our control problems in ASP and present our complete encoding
for Kemeny-CCAC. However, we mention here that although
ASP encodings for NP problems (and in fact even $\deltatwo$ problems) are fairly straightforward and common in the
literature, encoding problems in \sigmatwo\ typically requires more advanced
(and less intuitive) techniques
such as saturation~\cite{eiter:09}. 
We consider how this saturation technique can be used for our problems
and present an ASP encoding of Kemeny-CCAC
that uses saturation. %
We also discuss and compare
other encoding approaches for problems at this high level of complexity.

\subsection{Preliminaries on Answer Set Programming}

We briefly state the relevant definitions from ASP for our encoding. (See~Gebser et 
al.~\shortcite{geb-kam-kau-sch:b:answer-set-book} for more detailed 
definitions.) A {\em disjunctive logic program} is comprised of a finite set of 
rules 
of the form $a_1 \, | \dots | \, a_h \leftarrow b_1, b_2, \dots, b_m,\, {\rm not}\,b_{m+1}, \dots,\, {\rm 
not}\, b_n$ where each of $a_1,\dots,a_h, b_1, \dots, b_n$ are atoms and each atom is a 
constant or a predicate of the form $p(t_1, \dots, t_k)$ such that $k \ge 1$ and each $t_i$ 
is 
a constant or a variable. We indicate that $p$ is a $k$-ary predicate by writing
$p/k$.
The left side of the ``$\leftarrow$''
is the head of the rule and the right side is the body
of the rule.
In the rule, ``$|$'' denotes disjunction,
``$,$'' denotes conjunction, %
and ``not'' 
refers to default negation.
Uppercase characters are used to denote variables.
A {\em ground} program is a program that contains no variables.
A subset $S$ of the ground atoms satisfies a rule if
$\{b_1,\dots,b_m\}\subseteq S$ and
$\{b_{m+1},\dots,b_n\}\cap S=\emptyset$ implies
$a_i\in S$ for some $1 \leq i\leq h$.
A {\em model} is a subset of the ground atoms that satisfies each rule.
An {\em answer set} is a minimal model with respect to set inclusion.

A {\em fact} is a rule with no 
body and an {\em integrity constraint} is a rule with no head. So, a fact occurs 
in every answer set, and an integrity constraint eliminates answer sets where 
its body is satisfied.
We additionally use choice rules with cardinality 
constraints, which can be used to generate subsets of ground atoms within a 
given bound, and
aggregates such as {\tt \#count} and {\tt \#sum} that count/sum 
ground atoms in a statement.

It is customary to encode a decision problem as a logic program such
that the answer sets of this program correspond to certificates for
``yes'' answers to the problem. Under this approach, a ``no'' answer
is certified by the lack of an answer set. However, there are
circumstances in which a ``no'' answer must be certified by an answer
set (e.g., problems in \conp). %
The saturation technique (see, e.g., Eiter et al.~\shortcite{eiter:09}) achieves
this by designing a logic program that has a unique answer set
including a special token atom if and only if the answer to the
original decision problem is ``no.'' This answer set also contains the
set $S$ of all atoms that would be candidate certificates for ``yes''
answers. Rules are added so that every time the token atom is
generated, all atoms in $S$ are generated as well, thus ``saturating''
the model.
Informally, this %
allows us to encode a ``coNP check''
into our program, which along with an ``NP guess,'' allows us to encode
problems in \sigmatwo.

\subsection{Encoding Kemeny-CCAC in ASP}
We assume that the input to our problem is given as a list of facts. For 
\elec-CCAC,
our input consists of a fact for the number of registered candidates, the
number of unregistered candidates, the addition limit, the preferred candidate,
and facts that describe 
the voters. For the voters, we follow the approach used in Democratix~\cite{cha-pfa:c:democratix}, where each distinct vote $(c_1 >_i \dots >_i c_m)$
is represented by $m$ atoms of the form 
${\rm p}(i, j, c)$ meaning candidate $c$ is the $j$th-preferred candidate by 
vote $i$. The corresponding count is represented by ${\rm votecount}(i, k)$, 
meaning $k$ voters have vote~$i$.

We present our complete encoding for Kemeny-CCAC by presenting 
all of the rules of the encoding, split into ``guess,'' ``check,'' and
``saturate'' parts, and explaining the crucial aspects of each part.
Informally, the guess part will guess a subset of unregistered candidates
to add and a consensus such that the preferred candidate wins. The check part
(along with the saturate part) ensures that for the guessed set of added candidates,
no candidate beats the preferred candidate.

\begin{figure}[h]
\scriptsize %
\begin{align}
    & {\rm preference}(1..P) \leftarrow {\rm prefnum}(P).\\
    &\omit\rlap{\% Registered candidates.} \nonumber \\
    & {\rm candidate}(1..C) \leftarrow {\rm rcandnum}(C).\\
    &\omit\rlap{\% Unregistered candidates.} \nonumber \\
    & {\rm ucandidate}((M+1)..(M+N))\! \leftarrow\! {\rm rcandnum}(M),\! {\rm ucandnum}(N). \\
    &\omit\rlap{\% Guess a subset of at most $K$ candidates to add.} \nonumber \\
    &\{ {\rm candidate}(C) : {\rm ucandidate}(C) \} K \leftarrow {\rm limit}(K). \\
     &{\rm candnum}(N) \leftarrow N = {\tt \#count}\{{\rm candidate}(C) : {\rm candidate}(C)\}.\\
    &\omit\rlap{\% Number of times candidate $C$ is ranked below $D$.} \nonumber \\
    &{\rm wrank}(P,C,D)                          \leftarrow {\rm p}(P,X,C), {\rm p}(P,Y,D), Y < X. \label{ln:wrank} \\
    &{\rm wrankC}(C,D,N)                         \leftarrow {\rm candidate}(C), {\rm candidate}(D),\nonumber \\
    & \qquad \qquad N = {\tt \#sum}\{VC,P : {\rm votecount}(P,VC), {\rm wrank}(P,C,D)\}. \label{ln:wrankc} \\
    &{\rm position}(1..M)                       \leftarrow {\rm candnum}(M). \\
    &\omit\rlap{\% Guess a consensus.}\\
    &{\rm gpref}(X,C) \ |\ {\rm ungpref}(X,C) \leftarrow {\rm position}(X), {\rm candidate}(C).\\
    &                                             \leftarrow {\rm gpref}(X,C), {\rm gpref}(Y,C), X \neq Y.\\
    &                                             \leftarrow {\rm gpref}(X,C), {\rm gpref}(X,D), D \neq C.\\
    & \leftarrow {\rm grepf}(X,C), {\rm ungpref}(X,C).\\
    &\omit\rlap{\% Loop checks if all possible positions for a given cand.\ are in ${\rm ungpref}$.} \nonumber \\
    & {\rm npos}(X,Y) \leftarrow {\rm position}(X), Y = X+1.\\
    & {\rm countTo}(C,1) \leftarrow {\rm ungpref}(1,C).\\
    & {\rm countTo}(C,X) \leftarrow {\rm countTo}(C,Y), {\rm npos}(Y,X), {\rm ungpref}(X,C).\\
    & \leftarrow {\rm countTo}(C,X), {\rm candidate}(C), {\rm candnum}(X).\\
    &\omit\rlap{\% In the guessed consensus $C > D$.}\\
    &{\rm rank}(C,D)                          \leftarrow {\rm gpref}(X,C), {\rm gpref}(Y,D), X < Y. \label{ln:rank} \\
    &\omit\rlap{\% Number of votes that disagree on $C$ and $D$.}\\
    &{\rm gwrankC}(C,D,N)                      \leftarrow {\rm rank}(C,D), {\rm wrankC}(C,D,N). \label{ln:gwrankc} \\
    &                                             \leftarrow {\rm preferredCand}(X), {\rm gpref}(Y,X), {\rm position}(Y), Y \neq 1.
\end{align}
\caption{Rules of the guess part of Kemeny-CCAC.}
\label{fig:kemeny-ccac-guess}
\end{figure}

Figure~\ref{fig:kemeny-ccac-guess} shows guess part of
Kemeny-CCAC that
assumes an input as described
in Democratix~\cite{cha-pfa:c:democratix}, but extended with predicates
for the control problem.
We start by guessing (with
a choice rule) %
a subset of at most $K$ of the unregistered candidates
to add to the election
and we update the number of candidates (${\rm candnum}/1$).
We then define predicates
${\rm wrank}/3$ and ${\rm wrankC}/3$ to define the number of times that a
candidate is ranked ``worse'' than another in the given election, and we guess a
consensus. This generally follows what is done in
Democratix~\cite{cha-pfa:c:democratix}. Specifically, we follow the naming
conventions for different predicates, and 
rules~\ref{ln:wrank}, \ref{ln:wrankc}, \ref{ln:rank}, and~\ref{ln:gwrankc}
are from Democratix.
However, our rules to guess a consensus %
(${\rm gpref}/2$) are more involved, since we use a head disjunction.

\begin{figure}[h]
\scriptsize %
\begin{align}
    &\omit\rlap{\% Guess another consensus.} \nonumber \\
    &{\rm gpref}'(X,C)\ |\ {\rm ungpref}'(X,C) \leftarrow {\rm position}(X), {\rm candidate}(C).\\
    &{\rm sat}                                    \leftarrow {\rm gpref}'(X,C), {\rm gpref}'(Y,C), X \neq Y.\\
    &{\rm sat}                                    \leftarrow {\rm gpref}'(X,C), {\rm gpref}'(X,D), D \neq C.\\
    &\omit\rlap{\% Loop checks if all possible positions for a given cand.\ are in ${\rm ungpref}'$.} \nonumber \\
    &{\rm sat}                                    \leftarrow {\rm gpref}'(X,C), {\rm ungpref}'(X,C).\\
    &{\rm countTo}'(C,1)                    \leftarrow {\rm ungpref}'(1,C).\\
    &{\rm countTo}'(C,X)                    \leftarrow {\rm countTo}'(C,Y),\! {\rm npos}(Y,X),\! {\rm ungpref}'(X,\!C).\\
    &\omit\rlap{\% Saturate if all possible positions for a given candidate are in ${\rm ungpref}'$,} \nonumber \\
    &\omit\rlap{\% which means a candidate is not ranked in the guess.} \nonumber \\
    &{\rm sat}                                    \leftarrow {\rm countTo}'(C,X), {\rm candidate}(C), {\rm candnum}(X). \\
    &\omit\rlap{\% In the guessed consensus $C > D$.}\nonumber \\
    &{\rm rank}'(C,D)                              \leftarrow {\rm gpref}'(X,C), {\rm gpref}'(Y,D), X < Y.\\
    &\omit\rlap{\% Number of votes that disagree on $C$ and $D$.}\\
    &{\rm gwrankC}'(C,D,N)                         \leftarrow {\rm rank}'(C,D), {\rm wrankC}(C,D,N).\\
    &{\rm sat}\leftarrow {\tt \#sum}\{M,C1,C2,pos : {\rm gwrankC}'(C1,C2,M);\nonumber\\
    &\hspace{0.65in} {\rm -}N,D1,D2,neg : {\rm gwrankC}(D1,D2,N)\} >= 0. \label{ln:sum}\\
    &{\rm sat}\leftarrow {\rm preferredCand}(X), {\rm gpref}'(1,X).
\end{align}
\caption{Rules of the check part of Kemeny-CCAC.}
\label{fig:kemeny-ccac-check}
\end{figure}

Figure~\ref{fig:kemeny-ccac-check} shows the rules of the check
part of Kemeny-CCAC. The check part essentially checks that given the added candidates
from the guess, there is no candidate that beats %
the preferred candidate.
Note that this is quite similar to the guess part (Figure~\ref{fig:kemeny-ccac-guess}). However
what were integrity constraints there %
are now rules that generate the special
saturation token ${\rm sat}$.
We describe two important aspects of the check part below.

In the check part, we start by guessing %
a possible consensus. This is
done in the same way as the guess part of Kemeny-CCAC.
In line~\ref{ln:sum} we use a ${\tt \#sum}$ aggregate that generates the saturation
atom if a candidate other than the preferred candidate has a lower Kemeny score.
Note that the use of saturation-dependent aggregates may causes undesirable
behavior in the check part of an encoding and the interpretation of aggregates
is solver dependent. However, the interpretation of
aggregates in Clingo~4 (see Harrison et al.~\shortcite{har-lif-yan:c:gringo}) allows us to use
the ${\tt \#sum}$ aggregate
in this way. We mention that our use of aggregates here follows how they
are used in the saturation encodings in  Abseher et al.~\shortcite{abs-bli-cha-dus-wol:j:secure-sets}.

\begin{figure}[h]
\scriptsize
\begin{align}
    & {\rm gpref}'(X,C)                           \leftarrow {\rm position}(X), {\rm candidate}(C), {\rm sat}. \hspace{1.5cm}\\
    & {\rm ungpref}'(X,C)                        \leftarrow {\rm position}(X), {\rm candidate}(C), {\rm sat}.\\
    & {\rm possibleCount}(0..X)                    \leftarrow {\rm voternum}(X).\\
    & {\rm gwrankC}'(C,D,N)                        \leftarrow {\rm candidate}(C), {\rm candidate}(D),\nonumber \\
    & \hspace{4.75cm}{\rm possibleCount}(N), {\rm sat}. \\
    & {\rm rank}'(C,D)                              \leftarrow {\rm candidate}(C), {\rm candidate}(D), {\rm sat}.\\
    & {\rm countTo}'(C,N)                       \leftarrow {\rm candidate}(C), {\rm position}(N), {\rm sat}.\\
                                                & \leftarrow {\rm not}\ {\rm sat}.
\end{align}
\caption{Rules of the saturation part of Kemeny-CCAC.}
\label{fig:kemeny-ccac-sat}
\end{figure}

The final part of our encoding is the saturation step. Figure~\ref{fig:kemeny-ccac-sat}
shows the rules that ensure that an ``incorrect'' guess for the check program will
cause all of the predicates that depend on the guessed consensus ${\rm gpref}'/2$
to have all possible values in the stable model, thus saturating the solution.

Combining the guess, check, and saturate parts we obtain our complete
ASP program that is
satisfiable if and only if there is a subset of unregistered candidates that can
be added such that the preferred candidate is a winner.

The above encoding for Kemeny-CCAC is not as straightforward as ASP encodings for
problems at the NP level, but we still have a close relationship to the
definition of the problem and retain easy adaptability to minor changes in
the problem description. For example, it is not difficult to change the
above encoding to work for Kemeny-CCDC.

As with many
declarative problem solving tools, there is a trade-off between expressibility and
performance.
In our tests using
Clingo~4.5.4 and %
Preflib~\cite{mat-wal:c:preflib}
(the standard dataset of real-world preference data),
we noticed the above encoding
suffers from the so-called ``grounding bottleneck.'' 
That is, even on instances of Kemeny-CCAC
with around 10 candidates, we are faced with prohibitively
large programs after all variables are instantiated. 
This is in part due to the use of loops in
saturation (which is the standard way to replace default negation, the use of
which is limited in saturation), which were found in related work on encoding
\sigmatwo-complete problems
in ASP to have a strong negative impact on performance~\cite{gag-man-ron-wal-wol:j:improved-asp-argumentation}.
For a preliminary test of our encoding, we attempted to solve the Kemeny-CCAC problem for all
elections with 4 or more candidates having complete strict
order votes\footnote{\urlstyle{rm}\url{http://www.preflib.org/data/format.php#soc}}
making the first candidate the preferred candidate and holding out
the last fifth of the candidates as unregistered with an add limit
equal to one third of the number of unregistered candidates.
There were 215 elections
in total, and we were able to solve 114 of them using a timeout of 1
hour and a limit of 16GB of memory.
Some noteworthy
instances we  could solve
are summarized in Table~\ref{t:remarkable-instances}.
Of the 101 instances that we were not able to solve, 56 ran out of
memory and the rest timed out.
Despite its limitations, our %
encoding is an important
starting point for future optimization and comparison of techniques. In particular,
we are interested in how techniques for overcoming the grounding bottleneck
such as rewriting large rules~\cite{bic-mor-wol:j:nonground}
can improve the performance of our
encoding. %

\begin{table} %
\scriptsize
\centering
 \begin{tabular}{|c|c|c|c|c|c|}
 \hline
 Instance&\# Reg.&\# Unreg.&\# Voters&Seconds&Control Possible\\
 \hline
ED-9-2&5&2&153&1.79\dag&No\\
 \hline
ED-9-1&7&2&146&368.036\ddag&No\\
 \hline
ED-15-48&8&2&4&193.133\ddag&Yes\\
 \hline
ED-15-78&9&3&4&78.132\ddag&Yes\\
 \hline
 ED-6-4&11&3&9&3.619\dag&No\\
 \hline
 \end{tabular}
\caption{Details on some of the Preflib instances we were able to
  solve. The times reported were obtained on AMD Opteron(tm) 6180 SE
  (\dag) and AMD Opteron(tm) 6282 SE (\ddag) processors.}
\label{t:remarkable-instances}
\end{table}

\subsection{Related Encoding Approaches}
We presented an encoding of Kemeny-CCAC using the saturation technique
since it is the standard technique to encode \sigmatwo-complete problems in ASP
and a starting point for exploring how these problems can be encoded.
There are alternatives to our approach.

Eiter and Polleres~\shortcite{eiter:06} address
the issue of having to use advanced techniques, like saturation, when
writing ASP encodings of \sigmatwo\ problems
by providing a template
``meta-interpreter'' and a transformation of ASP programs 
that can, together, be used to integrate a program %
that guesses a solution 
to the \sigmatwo\ problem with a program
that checks whether the guessed
solution is incorrect. Their combination %
amounts to a ``guess and 
check'' encoding of a \sigmatwo\ problem.\footnote{Note this is different and much
more involved than
the guess and check technique for \np\ problems
from, e.g.,~Eiter et al.~\shortcite{eiter:09}.}
However, this approach does %
not support ASP programs %
with aggregates (e.g., {\tt \#count}),
which leads to more complex and somewhat less intuitive programs.
To work around this issue,
one could use the {\tt lp2normal} tool~\cite{bomanson:14}, which transforms ASP
programs with aggregates into equivalent ASP programs without aggregates.

The work by Gebser et al.~\shortcite{geb-kam-sch:j:complex-optimization-asp} also
addresses ASP encodings for problems with complexity higher than
\np\ with a different metaprogramming
approach that does support aggregates by using optimization rules.

The framework of \emph{stable-unstable} semantics by Bogaerts et al.~\shortcite{bog-jan-tas:j:stable-unstable} provides a similar ``guess and check'' strategy for solving problems in $\sigmatwo$. They point out that an advantage of the stable-unstable semantics is they can be easily extended to represent problems at any level of the polynomial hierarchy. In their implementation guess and check programs are each normalized, essentially discarding aggregates altogether.

It is an interesting direction for future work to determine how the performance of the above techniques and alternative paradigms for constraint satisfaction
compare with
saturation encodings for very hard control problems in terms of implementation
and adaptability. %

\section{Future Work}\label{sec:conclusion}
In addition to the future work on ASP described at the end of the previous section, it will be
interesting to see if our newly-defined simple \sigmatwo-complete
problems will be useful in proving other problems \sigmatwo-hard, in particular
the remaining control cases and other election-attack problems such as manipulation
for systems with hard winner problems.

\smallskip

\noindent
{\bf Acknowledgments:} 
We thank Andreas Pfandler for helpful conversations and
we thank the referees for their helpful comments and suggestions.
This work was supported in part by an
NSF Graduate
Research Fellowship under grant no.\
DGE-1102937.

{
\fontsize{9.0pt}{10.0pt}
 \selectfont
\bibliography{gryww,pepper}
\bibliographystyle{abbrv} %
}

\appendix

\section{Appendix} 

\subsection{Kemeny}

\subsubsection{Proof of Theorem~\ref{thm:kem:ccdcstar}:
Kemeny-\ccdcstar\ is $\sigmatwo$-complete}
\label{a:thm:kem:ccdcstar}

Kemeny-Score
was shown to be \np-hard by a reduction from Feedback-Arc-Set by
Bartholdi, Tovey, and Trick~\shortcite{bar-tov-tri:j:who-won}, which in
turn was shown to be \np-hard by Karp~\shortcite{kar:b:reducibilities}.
The complete proof of the \np-hardness of Kemeny-Score 
can be viewed as consisting of the following three steps.
\begin{enumerate}
\item Vertex-Cover is \np-hard~\cite{kar:b:reducibilities}.
\item Vertex-Cover $\leq_m^p$ Feedback-Arc-Set~\cite{kar:b:reducibilities}.
\item Feedback-Arc-Set $\leq_m^p$
Kemeny-Score~\cite{bar-tov-tri:j:who-won}.
\end{enumerate}

Kemeny-Winner was shown to be $\thetatwo$-hard by a chain
of three reductions, basically a ``lifted'' version of
the \np-hardness proof for Kemeny-Score.
For the lifted-to-$\thetatwo$ version, Hemaspaandra, Spakowski, and
Vogel~\shortcite{hem-spa-vog:j:kemeny} define suitable $\thetatwo$-complete
equivalent problems, and show the following.\footnote{Fitzsimmons
and Hemaspaandra~\shortcite{fit-hem:c:high-multiplicity} recently showed a similar
``lifting'' to $\deltatwo$, to show that Kemeny-Winner for weighted elections
(elections where each voter has a corresponding integral weight) and for
high-multiplicity %
elections (where the voters are represented as a list of distinct
votes and their corresponding counts) is hard for~\deltatwo.}
\begin{enumerate}
\item Vertex-Cover-Member is $\thetatwo$-hard. 
\item Vertex-Cover-Member $\leq_m^p$ Feedback-Arc-Set-Member.
\item Feedback-Arc-Set-Member $\leq_m^p$ Kemeny-Winner.
\end{enumerate}

In order to show that Kemeny-\ccdcstar\ is $\sigmatwo$-hard,
we lift the reductions to $\sigmatwo$.
We already defined an appropriate 
$\sigmatwo$-complete analogue of Vertex-Cover
and below we define an appropriate 
$\sigmatwo$-complete analogue of 
Feedback-Arc-Set, and we show the following.
\begin{enumerate}
\item Vertex-Cover-Member-Select is $\sigmatwo$-hard (Lemma~\ref{l:vcms}).
\item Vertex-Cover-Member-Select $\leq_m^p$ Feedback-Arc-Set-Member-Select
(Lemma~\ref{l:a2}).
\item Feedback-Arc-Set-Member-Select $\leq_m^p$  Kemeny-\ccdcstar\
(Lemma~\ref{l:a3}).
\end{enumerate}

The $\sigmatwo$-complete analogue of
Feedback-Arc-Set is defined as follows.

\prob{Feedback-Arc-Set-Member-Select}
{Irreflexive and antisymmetric directed graph $G = (V,A)$,
a set $V' \subseteq V - \{\hat{v}\}$ of deletable vertices,
delete limit $k$, and vertex $\hat{v} \in V$.}
{Does there exist a set $W \subseteq V'$ of at most $k$ deletable vertices
such that some minimum size feedback arc set of $G - W$ contains
all arcs entering $\hat{v}$?}

The second and third parts of the lifting use similar constructions
as in the \np, $\thetatwo$, and $\deltatwo$ cases.
As we saw previously in the proof Lemma~\ref{l:vcms},
the proof
that Vertex-Cover-Member-Select is $\sigmatwo$-hard requires significantly
more work.

\begin{lemma}
\label{l:a2}
Vertex-Cover-Member-Select $\leq_m^p$
Feedback-Arc-Set-Member-Select.
\end{lemma}

\begin{proofs}
Given a graph $G$, define digraph $f(G) = (\widehat{V},A)$
as in the standard reduction 
from Vertex-Cover to Feedback-Arc-Set from~\cite{kar:b:reducibilities}.

\begin{enumerate}
\item $\widehat{V} = \{v,v' \ | \ v \in V\}$.
\item $A = \{(v,v') \ | \ v \in V\} \cup 
\{(v',w), (w',v) \ | \ \{v,w\} \in E\}$.
\end{enumerate}

We know from~\cite[Lemma 4.8]{hem-spa-vog:j:kemeny} that for every graph $G'$ 
and vertex $\hat{v}$ in $G'$,
$G'$ has a minimum size vertex cover containing $\hat{v}$
if and only if $f(G')$ has a minimum size feedback arc set containing
$(\hat{v},\hat{v}')$, which is the only arc entering $\hat{v}'$.

Our reduction from Vertex-Cover-Member-Select to
Feedback-Arc-Set-Member-Select
maps $(G,V',k,\hat{v})$ to $(H,V' - \{\hat{v}\},k,\hat{v}')$,
where $H = f(G)$.
Let $W \subseteq V' - \{\hat{v}\}$.
Then from~\cite[Lemma 4.8]{hem-spa-vog:j:kemeny},
$G-W$ has a minimum size vertex cover containing $\hat{v}$
if and only if $f(G-W)$ has a minimum size feedback arc set containing
$(\hat{v},\hat{v}')$, which is the only arc entering $\hat{v}'$.

Now consider $H$. Note that if we delete a vertex $v$, we do not have any
cycles going through $v'$ and so $A'$ is a minimum feedback arc set of
$H - W$ if and only if $A'$ is a minimum feedback arc set of 
$H - W - \{v' \ | \ v \in W\}$.
Since $H - W - \{v' \ | \ v \in W\} = f(G-W)$, it follows that
$G-W$ has a minimum size vertex cover containing $\hat{v}$
if and only if $H-W$ has a minimum size feedback arc set containing
$(\hat{v},\hat{v}')$, which is the only arc entering $\hat{v}'$.
\end{proofs}

\begin{lemma}
\label{l:a3}
Feedback-Arc-Set-Member-Select $\leq_m^p$ Kemeny-CCDC*.
\end{lemma}

\begin{proofs}
We use the construction from~\cite{bar-tov-tri:j:who-won}.
Given an irreflexive and antisymmetric digraph
$G = (C,A)$, let $C$ be the set of candidates and use McGarvey's
construction~\cite{mcg:j:election-graph} 
to construct in polynomial time a set of voters such that
for every arc $(v,w)$ in $A$, there are exactly two more voters who
prefer $v$ to $w$ than who prefer $w$ to $v$ and if there are no
arcs between $v$ and $w$ then the same number of voters prefer
$v$ to $w$ as $w$ to $v$.
From the proof of Lemma 4.2 of~\cite{hem-spa-vog:j:kemeny},
it holds that for every $C' \subseteq C$ and
for every $c \in C'$ that
some minimum feedback arc set of $(C',A \cap (C' \times C'))$ contains
all arcs entering $c$
if and only if $c$ is a Kemeny winner of $(C',V)$. This then
shows that Feedback-Arc-Set-Member-Select reduces to
Kemeny-CCDC*.
\end{proofs}

This completes the proof that Kemeny-\ccdcstar\ is \sigmatwo-complete.

\medskip

Do we always get this jump from a $\thetatwo$-complete winner problem
to a $\sigmatwo$-complete control problem?
Dwork et al.~\shortcite{dwo-kum-nao-siv:c:rank-aggregation}
show that Kemeny-Winner is already \np-hard if we have four voters.
And Fitzsimmons and Hemaspaandra~\shortcite{fit-hem:c:high-multiplicity} show
that Kemeny-Winner for four voters is still $\thetatwo$-complete.
Certainly the voter control cases for Kemeny with four voters are still in
$\thetatwo$, and the control by adding voters and control by deleting voters
cases are clearly $\thetatwo$-complete.
What about the candidate control cases? 

\begin{theorem}
Kemeny-\ccdcstar\ is $\sigmatwo$-complete, even for four voters.
\end{theorem}

\begin{proofs}
To show hardness, we argue as in
the proof that Kemeny-Winner for four voters is 
$\thetatwo$-hard from~\cite{fit-hem:c:high-multiplicity}.

Given an irreflexive and antisymmetric digraph $G = (V,A)$ and vertex
$\hat{v} \in V$,
we first compute an irreflexive and antisymmetric digraph $\widehat{G}$ as done
in~\cite{dwo-kum-nao-siv:c:rank-aggregation}.
$\widehat{G} = (\widehat{V},\widehat{A})$ such that $\widehat{V} = V \cup A$ and
$\widehat{A} = \{(v,(v,w)), ((v,w),w) \ | \ (v,w) \in A\}$.
Let $W \subseteq V - \{\hat{v}\}$.
Note that $G-W$ has a feedback arc set of size $\ell$
that contains all arcs entering $\hat{v}$ if and only if
$\widehat{G}-W$ has a feedback arc set of size $\ell$ that contains
all arcs entering $\hat{v}$.
It follows that
$(G,V',k,\hat{v})$ is in Feedback-Arc-Set-Member-Select
if and only if
$(\widehat{G},V',k,\hat{v})$ is in Feedback-Arc-Set-Member-Select.
Now apply the 
reduction from Feedback-Arc-Set-Member-Select to
Kemeny-\ccdcstar\ from Lemma~\ref{l:a3} to
$(\widehat{G},V',k,\hat{v})$.  The construction
from Dwork et al.~\shortcite{dwo-kum-nao-siv:c:rank-aggregation} shows
that we need only four voters in the election.~\end{proofs}

\subsubsection{Proof of Theorem~\ref{thm:kem:ccac}:
Kemeny-CCAC is $\sigmatwo$-complete}
\label{a:thm:kem:ccac}

Membership follows from Corollary~\ref{cor:upperbound}.
The proof of hardness is similar to the proof of hardness
for \ccdcstar: We define suitable $\sigmatwo$-complete
versions of Vertex-Cover and Feedback-Arc-Set and show the following.

\begin{enumerate}
\item Vertex-Cover-Member-Add is $\sigmatwo$-hard.
\item Vertex-Cover-Member-Add $\leq_m^p$ Feedback-Arc-Set-Member-Add.
\item Feedback-Arc-Set-Member-Add $\leq_m^p$  Kemeny-CCAC.
\end{enumerate}

\prob{Vertex-Cover-Member-Add}
{Graph $G = (V \cup V',E)$, set of addable vertices $V'$, 
addition limit $k$, and vertex $\hat{v} \in V$.}
{Does there exist a set $W \subseteq V'$ of at most $k$ addable vertices
such that
$\hat{v}$ is a member of a minimum vertex cover of
$(V \cup W,E)$?\footnote{We slightly abuse notation by
writing $(V \cup W,E)$ instead of
$(V \cup W,E')$, where $E'$ is the restriction of $E$ to 
$V \cup W$.}}

\prob{Feedback-Arc-Set-Member-Add}
{Irreflexive and antisymmetric directed graph $G = (V \cup V',A)$,
a set $V'$ of addable vertices,
addition limit $k$, and vertex $\hat{v} \in V$.}
{Does there exist a set $W \subseteq V'$ of at most $k$ addable vertices
such that some minimum size feedback arc set of $(V \cup W,A)$
contains all arcs entering $\hat{v}$?}

The proof that
Vertex-Cover-Member-Add reduces to 
Feedback-Arc-Set-Member-Add is similar to the proof of Lemma~\ref{l:a2}
and the proof that Feedback-Arc-Set-Member-Add reduces
to Kemeny-CCAC is similar to the proof of Lemma~\ref{l:a3}.
This leaves the following lemma to prove.

\begin{lemma}
Vertex-Cover-Member-Add is $\sigmatwo$-complete.
\end{lemma}

\begin{proofs}
The upper bound is immediate.
For the lower bound, we will again reduce from QSAT$_2$.
We will modify the construction of the proof of Lemma~\ref{l:vcms},
which showed $\sigmatwo$-hardness for
Vertex-Cover-Member-Select.
Let $H$ be the graph from the proof of Lemma~\ref{l:vcms}.
We add an addable vertex $x'_i$ connected only to $x_i$ and
an addable vertex $\overline{x_i}'$ connected only to
$\overline{x_i}$. 
So, $V = V(H)$, $V' = \{x'_i, \overline{x_i}' \ | \ 1 \leq i \leq n\}$, 
and $E = E(H) \cup \{\{x_i, x'_i\}, \{\overline{x_i}, \overline{x_i}'\} \ | \
1 \leq i \leq n\}$.\footnote{We can not simply use $H$ with
the $x$-vertices as addable vertices, since in that case
$d$ would be a member of a minimum vertex cover after adding 
any nonempty set of vertices that corresponds to an assignment.}

Let $W \subseteq V'$.  Let $H' = (V \cup W,E)$.
Some relevant properties of $H'$ are as follows.

\begin{enumerate}
\item \label{one}If $X$ is a vertex cover, then $X$ is still a vertex
cover if we replace $x'_i$ by $x_i$ and 
$\overline{x_i}'$ by $\overline{x_i}$. In particular, this
implies that there is a minimum vertex cover that does
not contain any vertex in $W$. 
\item \label{two} Every vertex cover of $H'$ contains at least one
of $\{x_i,\overline{x_i}\}$, at least one of $\{y_i,\overline{y_i}\}$,
at least three of $\{a_j, b_j,c_j,d_j\}$,
at least one of $\{x_i, x'_i\}$ for $x'_i \in W$, and
at least one of $\{\overline{x_i},\overline{x_i}\}$ for
$\overline{x_i}' \in W$. 
\item \label{four} If $X$ is a vertex cover of size $2n + 3m$, then
$\hat{v} \not \in X$ and $V' \cap X = \emptyset$ and 
$X \cap \{x_i, \overline{x_i}, y_i, \overline{y_i}  \ | \ 1 \leq i \leq n\}$
corresponds to a satisfying assignment for $\phi$.
\item \label{five} If $\alpha$ is a satisfying assignment for $\phi$, then
there is a vertex cover $X$ of size $2n+3m$ of $(V,E)$ such that
$X \cap \{x_i, \overline{x_i}, y_i, \overline{y_i} \ | \ 1 \leq i \leq n\}$
corresponds to this assignment.
\item \label{six} If for all $i, 1 \leq i \leq n$, 
$\{x_i',\overline{x_i}'\} \not \subseteq W$, then 
there is a vertex cover of size 
$2n + 3m + 1$ that includes $\hat{v}$ and that
does not contain any vertex in $W$. 
\end{enumerate}

We will show that $\exists x_1 \cdots \exists x_n
\neg (\exists y_1 \cdots \exists y_n \phi(x_1, \ldots, x_n,\\ y_1, \ldots, y_n))$
if and only if there exists a set $W \subseteq V'$ 
of at most $n$ addable vertices 
such that 
$\hat{v}$ is a member of a minimum vertex cover of 
$(V \cup W,E)$.

First assume that $\exists x_1 \cdots \exists x_n
\neg (\exists y_1 \cdots \exists y_n \phi(x_1, \ldots, x_n,\\ y_1, \ldots, y_n))$.
Let $\alpha \in \{0,1\}^n$ be an assignment to $x_1 \cdots x_n$ such that
$\phi(\alpha_1, \ldots, \alpha_n, y_1, \ldots, y_n)$ is not 
satisfiable.
Let $W$ be the set of vertices in $V'$ corresponding to $\alpha$, i.e.,
$W = \{x'_i \ | \  \alpha_i = 1\} \cup \{\overline{x_i}' \ | \ \alpha_i = 0\}$.
Then $(V \cup W,E)$ does not have a vertex cover of size $2n + 3m$
(by~\ref{four}),
and $(V \cup W,E)$ does have a vertex cover of size $2n + 3m + 1$ that includes $\hat{v}$ (by~\ref{six}).

For the converse, suppose that $W$ is a set of at most $n$ vertices from
$\{x'_i, \overline{x_i}' \ | \ 1 \leq i \leq n\}$ such that
$\hat{v}$ is a member of a minimum vertex cover of $(V \cup W,E)$.
If $W$ corresponds to an assignment $\alpha$, it is easy to see
that $\phi(\alpha_1, \ldots, \alpha_n, y_1, \ldots, y_n)$ is not 
satisfiable. But $W$ does not necessarily correspond to an assignment,
since it is possible for $W$ to contain neither or both of
$\{x_i',\overline{x_i}'\}$.
Let $\widehat{n}$ be the number of 
$\{x_i', \overline{x_i}'\}$ pairs that are unadded, i.e.,
for which $\{x_i', \overline{x_i}'\} \cap W = \emptyset$.
Note (by~\ref{two}) that the size of a vertex cover of $(V \cup W,E)$ is 
at least $\|W\| + \widehat{n} + n + 3m$, and that vertex covers of that
size do not include $\hat{v}$. Since $\hat{v}$ is a member of a minimum size
vertex cover of $(V\cup W,E)$, it follows that
$(V\cup W,E)$ does not have a vertex cover of size
$\|W\| + \widehat{n} + n + 3m$.

Let $\alpha \in \{0,1\}^n$ be an assignment to $x_1 \cdots x_n$ that is
consistent with $W$, in the sense that
for all $i, 1 \leq i \leq n$,
if $W \cap \{x'_i,\overline{x_i}'\} = \{x'_i\}$, then $\alpha_i = 1$ and
if $W \cap \{x'_i,\overline{x_i}'\} = \{\overline{x_i}'\}$, then $\alpha_i = 0$.
To complete the proof, suppose for a contradiction that
$\phi(\alpha_1, \ldots, \alpha_n, y_1, \ldots, y_n)$ is 
satisfiable.
Then (by~\ref{five}) $(V,E)$ has a vertex cover $X$ of size
$2n+3m$ such that $X \cap 
\{x_i, \overline{x_i} \ | \ 1 \leq i \leq n\}$ corresponds
to $\alpha$.
Clearly, $X \cup \{x_i \ | \ x_i' \in W\}
\cup \{\overline{x_i} \ | \ \overline{x_i}' \in W\}$ is a
vertex cover of $(V \cup W, E)$. 
Since $\|X \cap (\{x_i \ | \ x_i' \in W\} \cup
\{\overline{x_i} \ | \ \overline{x_i}' \in W\})\| = n - \widehat{n}$,
the size of this vertex cover is $\|W\| + \|X\| - (n + \widehat{n}) =
\|W\| + \widehat{n} + n + 3m$, which
implies that $(V\cup W,E)$ has a vertex cover of size 
$\|W\| + \widehat{n} + n + 3m$, which is a contradiction.~\end{proofs}

\subsubsection{Proof of Theorem~\ref{thm:kemprime:ccav}:
Kemeny$'$-CCAV is $\sigmatwo$-complete}

Note that in  the reduction from feedback arc set problems to
Kemeny control problems, vertices correspond to candidates and 
arcs roughly correspond to voters.
Also note that arc $(v,v')$ in Feedback-Arc-Set-Member-Add basically
correspond to vertex $v$. And so we can easily define
a version of Feedback-Arc-Set-Member-Add
where we add arcs instead of vertices. 

\prob{Feedback-Arc-Set-Member-Add-Arcs}
{Irreflexive and antisymmetric directed graph $G = (V,A \cup B)$,
a set $B$ of addable arcs,
addition limit $k$, and vertex $\hat{v} \in V$.}
{Does there exist a set $B' \subseteq B$ of at most $k$ addable arcs
such that some minimum size feedback arc set of $(V,A \cup B')$
contains all arcs entering $\hat{v}$?}

It is easy to see that
Vertex-Cover-Member-Add $\leq_m^p$ Feedback-Arc-Set-Member-Add-Arcs:
Given a graph $G = (V \cup W, E)$, define digraph
$f(G) = (\widehat{V},A \cup B)$ 
as in the standard reduction 
from Vertex-Cover to Feedback-Arc-Set from~\cite{kar:b:reducibilities},
with the set of addable arcs $B$ being the arcs corresponding to
the addable vertices $W$.

\begin{enumerate}
\item $\widehat{V} = \{v,v' \ | \ v \in V \cup W\}$.
\item $A = \{(v,v') \ | \ v \in V\} \cup
\{(v',w), (w',v) \ | \ \{v,w\} \in E\}$.
\item $B = \{(v,v') \ | \ v \in W\}$.
\end{enumerate}

Our reduction from Vertex-Cover-Member-Add to 
Feedback-Arc-Set-Member-Add-Arcs maps
$(G,W,k,\hat{v})$ to $((\widehat{V},A\cup B),B,k,\hat{v}')$.

As mentioned previously,
in the reduction from feedback arc set problems to
Kemeny control problems, vertices correspond to candidates and 
arcs roughly correspond to voters. If the voters vote with total
orders, each arc corresponds to two voters: arc $a \rightarrow b$ corresponds
to a voter voting $a > b > C - \{a,b\}$ and a voter voting
($C - \{a,b\})^r > a > b$~\cite{mcg:j:election-graph}.

There exist variations of Kemeny where the voters do not necessarily
list all of the candidates in their votes.
Dwork et al.~\cite{dwo-kum-nao-siv:c:rank-aggregation} consider a natural
variant of Kemeny, which we here call Kemeny$'$, 
where unlisted candidates in a vote do not contribute
to the distance to the Kemeny consensus and so do not increase the
Kemeny score. (The Kemeny consensus
is still a complete total order.) So, in this model, a voter
voting $a > b$ corresponds exactly to an arc $a \rightarrow b$. This
gives a straightforward reduction from
Feedback-Arc-Set-Member-Add-Arcs to Kemeny$'$-CCAV.

\subsection{Young}

\subsection{Proof of Theorem~\ref{thm:ismd}}\label{app:young-is}

\ismdproof

\subsubsection{Proof of Theorem~\ref{thm:young:ccdv}}\label{app:young}

\youngccdv

\subsubsection{Details of the proof of Theorem~\ref{thm:young:ccdv}}\label{app:young2}

Given graph $G = (V,E)$, a delete limit $k \leq \|V\|$, and a vertex
$\hat{v} \in V$, such that
$G$ has no isolated vertices and for every set $W$ of at most
$k$ vertices, $\alpha(G-W) \geq 3$,
we will show that $((V,E),k,\hat{v})$ is a positive instance
of Independent-Set-Member-Delete if and
only if we can make the Young score
of $p$ at least 2 and at least as high as the Young score of $q$
by deleting at most $k$ voters.

Let $W \subseteq V$ be a set of at most $k$ vertices
such that $\hat{v}$ is in a maximum independent set of $G - W$.
Delete the voters corresponding to $W$. We will show that
the Young score of $p$ is at least 2 and at least as high 
as the Young score of $q$.
Let $\widehat{W}$ be a maximum independent set of $G - W$ such that
$\hat{v} \in \widehat{W}$.  Then all the voters corresponding to
$\widehat{W}$ are of Type IA. It is easy to see that $p$ and $q$
are weak Condorcet winners in the set of voters that consists of
the voters corresponding to $\widehat{W}$, the voter of Type II, and
$\|\widehat{W}\| - 1$ voters of Type III. It follows that the Young scores 
of $p$ and $q$ are at least $2\alpha(G-W)$. It also follows from
the argument from~\cite[Lemma 2.4]{rot-spa-vog:j:young} 
that the Young score of $q$ is at most $2\alpha(G-W)$. 

For the converse, suppose $X$ is a set of at most $k$ voters such
that after deletion, $p$'s Young score is at least 2 and at least
as high as $q$'s Young score.  Then $X$ does not contain the
voter of Type II.  Let $W$ be the set of vertices corresponding
to the Type-I voters in $X$. Then the Young score of
$q$ is $2 \alpha(G - W)$. Let $\widehat{X}$ be a set of voters of
size $2 \alpha(G - W)$ such that $p$ is a weak Condorcet winner of $\widehat{X}$.
Without loss of generality, assume that $\widehat{X}$ does not contain voters
of Type IB (if it does, we can replace such a voter by a voter of 
Type III). In order for $p$ to be a weak Condorcet winner, the
vertices corresponding to the voters of Type I in $\widehat{X}$ form an 
independent set of $G - W$. Since all voters in $\widehat{X}$ of Type I are of Type IA, 
this independent set contains no vertices connected to $\hat{v}$ and
so we can safely add $\hat{v}$ to this independent set, giving us
an independent set containing $\hat{v}$ of size at least $\alpha(G-W)$.

\subsection{Dodgson Elections}\label{sec:dodgson}

In this section we consider the complexity of control for
Dodgson elections~\cite{dod:unpubMAYBE-without-embedded-citations:dodgson-voting-system}.
Recall that the Dodgson score of a candidate $c$ using the Dodgson election system is
the minimum number of swaps between adjacent candidates in votes such that $c$ is a Condorcet
winner and that the candidate(s) with the minimum score are the winner(s) of the
Dodgson election.
We will use the notation DodgsonScore$(c)$ to denote the Dodgson score of a candidate
$c$ in a given election.
As we did with Kemeny, we first consider the complexity of \ccdcstar.

\begin{theorem}\label{dod:ccdcstar}
Dodgson-\ccdcstar\ is \sigmatwo-complete.
\end{theorem}

\begin{proofs}
In our hardness proofs for control for Kemeny and Young elections, our approach was
to define a new simple \sigmatwo-complete analogue of Vertex-Cover or Independent-Set (the problems used to show NP-hardness for the score problems)
to then reduce to a control problem. 
Dodgson score was originally shown \np-hard by a reduction from
Exact-Cover-by-3-Sets~\cite{bar-tov-tri:j:who-won}.
\cite{hem-hem-rot:j:dodgson} provides
a reduction from Three-Dimensional-Matching (3DM) as part of the proof of 
\thetatwo-completeness of the winner problem.
There does exist a simple 
\sigmatwo-complete analogue of 3DM~\cite{mcl:j:radius}.
However, this analogue does not straightforwardly reduce to Dodgson
control problems.

Fortunately, in Dodgson there is a lot of
flexibility in how to construct the voters
in a reduction, and so we will directly reduce QSAT$_2$ to
Dodgson-\ccdcstar, though the construction is quite involved.
To better understand our reduction, it helps to first consider
a reduction from 3SAT to the Dodgson score problem, in which we ask if the Dodgson score
of a distinguished candidate is at most $k$.

Let $\phi(z_1,\ldots,z_n)$ be a Boolean formula in 3cnf, where
$\phi = \psi_1 \wedge \psi_2 \wedge \cdots \wedge \psi_m$ and
for each $i, 1 \le i \le m$, $\psi_i = \ell_{i,1} \vee \ell_{i,2} \vee \ell_{i,3}$.
Without loss of generality, let $n \ge 1$ and $m \ge 1$.
We now construct an instance of the Dodgson score problem.
An example of this construction is given in Example~\ref{ex:dod-3sat} below.
We begin by
showing the core parts of the construction (Blocks~I and~II).

Let $C = \{\widehat{z}_1, \ldots, \widehat{z}_n\} \cup
\{c_1, \ldots, c_m\} \cup \{c_{i,1}, c_{i,2}, c_{i,3}, \dots, c_{m,1}, c_{m,2}, c_{m,3}\} \cup \{q,b\}$.
Let there be the following voters. Note that ``$\cdots$'' in a vote denotes that the remaining
candidates are strictly ranked in a arbitrary, fixed, and easy to compute order.
Similarly, a set in a vote denotes that the
candidates in that set are strictly ranked with respect to that order.

\begin{description}

\item[Block I] For each $i, 1 \le i \le m$ and $j, 1 \le j \le 3$,
    \begin{itemize}
        \item One voter voting: $(c_i > c_{i,j} > q > \cdots)$.
    \end{itemize}
\item[Block II] For each $t, 1 \le t \le n$,
    \begin{itemize}
      \item One voter voting: $(\widehat{z}_t > \{c_{i,j}\ |\ \ell_{i,j} = z_t\} > q > \cdots)$.
      \item One voter voting: $(\widehat{z}_t > \{c_{i,j}\ |\ \ell_{i,j} = \overline{z_t}\} > q > \cdots)$.
    \end{itemize}
\end{description}
The above election can be easily padded so that the following properties hold.
\begin{itemize}
    \item $q$ needs one vote over each $c_i$, each $c_{i,j}$, and each $\widehat{z}_i$,
    and no votes over $b$.

    \item One swap in votes other than the Block~I and Block~II votes does not give $q$ a
        vote over any of the $c_i$, $c_{i,j}$, or $\widehat{z}_i$ candidates, i.e., one swap in votes other than
    Block~I and Block~II votes is useless.
    \end{itemize}

Note that for each $c_{i,j}$ candidate, $2n+3m-2$ (of the $2n + 3m$)
voters in Blocks~I and~II prefer $q$ to $c_{i,j}$,
that for each $\widehat{z}_i$ candidate, 
$2n+3m-2$ voters in Blocks~I and~II prefer $q$ to
$\widehat{z}_i$, and that  for each
$c_i$ candidate, $2n+3m-3$ of the
voters in Blocks~I and~II prefer $q$ to $c_i$.  We can ensure that $q$ needs one vote
over each of these candidates (by making $q$ tie pairwise with each
of these candidates) by adding the following $2n+3m-4$ voters.
We use the buffer candidate ``$b$'' to ensure that one swap outside of Blocks~I and~II
does not give $q$ a vote over any of the $c_i$, $c_{i,j}$, or $\widehat{z}_i$ candidates.
\begin{description}
    \item[Block III] \hfill
        \begin{itemize}
            \item One voter voting: $(\cdots > b > q > \{c_1, \ldots, c_m\})$.
            \item $2n+3m-5$ voters voting: $(\cdots > b > q)$.
        \end{itemize}
\end{description}

We will show in Lemma~\ref{lem:dodgsonscore} that $\phi$ is satisfiable if and
only if DodgsonScore$(q) \le 4m+n$.

Before proving the correctness of this construction, let's first consider the
following example, which will show how assignments to variables correspond to
swaps between adjacent candidates.

\begin{example}\label{ex:dod-3sat} 
Given the formula $\phi(z_1,z_2) = (z_1 \vee \overline{z_2} \vee z_1) \wedge (\overline{z_2} \vee \overline{z_1}
\vee z_2)$, our construction gives the following election.

$C = \{c_1, c_2, c_{1,1}, c_{1,2}, c_{1,3}, c_{2,1}, c_{2,2}, c_{2,3}, \widehat{z}_1, \widehat{z}_2, q, b\}$,
and we have the following voters. (We do not name the voters in Block~III since we
do not need to swap $q$ in these votes.)

\begin{description}

\item[Block~I] \hfill
\begin{itemize}
	\item $v_1$ voting: $(c_1 > c_{1,1} > q > \cdots).$ %
	\item $v_2$ voting: $(c_1 > c_{1,2} > q > \cdots).$ %
	\item $v_3$ voting: $(c_1 > c_{1,3} > q > \cdots).$ %
	\item $v_4$ voting: $(c_1 > c_{2,1} > q > \cdots).$ %
	\item $v_5$ voting: $(c_1 > c_{2,2} > q > \cdots).$ %
	\item $v_6$ voting: $(c_1 > c_{2,3} > q > \cdots).$ %
\end{itemize}
\item[Block~II] \hfill
\begin{itemize}
	\item $v_7$ voting: $(\widehat{z}_1 > c_{1,1} > c_{1,3} > q > \cdots).$ %
	\item $v_8$ voting: $(\widehat{z}_1 > c_{2,2} > q > \cdots).$ %
	\item $v_9$ voting: $(\widehat{z}_2 > c_{2,3} > q > \cdots).$ %
	\item $v_{10}$ voting: $(\widehat{z}_2 > c_{1,2} > c_{2,1} > q > \cdots).$ %
\end{itemize}

\item[Block~III] \hfill
\begin{itemize}
    \item One voter voting: $(\cdots > b > q > c_1 > c_2)$. %
    \item Five voters voting: $(\cdots > b > q)$. %
\end{itemize}
\end{description}

It is easy to see that the assignment $z_1=1,z_2=0$ is a satisfying assignment for $\phi$.
We will now show that this implies
that DodgsonScore$(q) \le 10$.

Swapping $q$ over a $c_{i,j}$ candidate in its Block~I vote will correspond to that clause literal
being true and swapping $q$ over a $c_{i,j}$ candidate in its Block~II vote will correspond to that clause literal being false.
Since $z_1$ is true, swap $q$ with $c_{1,1}$ in $v_1$ and $c_{1,3}$ in $v_3$ (these correspond to positive
$z_1$-literals in $\phi$) and then in Block~II swap $q$ with $c_{2,2}$ and $\widehat{z}_1$ in $v_8$ (this vote corresponds to the
negated $z_1$-literals). Similarly, since $z_2$ is false, swap $q$ with $c_{1,2}$ in $v_2$, swap $q$ with $c_{2,1}$ in $v_4$,
and then in Block~II swap $q$ with $c_{2,3}$ and $\widehat{z}_2$ in $v_{9}$. This takes eight swaps for $q$ to gain one vote over each of
the $\widehat{z}_i$ and $c_{i,j}$ candidates.

Note that since $\phi$ is satisfiable, for each clause there exists at least one true literal, and we just swapped $q$ over
the true clause literals
in Block~I. So, to gain one vote over $c_1$, swap $q$ with $c_1$ in $v_1$ (we already swapped
$q$ with $c_{1,1}$) and to gain one vote over $c_2$, swap $q$ with $c_2$ in $v_2$ (we already swapped $q$ with
    $c_{1,2}$). This takes exactly two swaps. Thus the Dodgson score of $q$ is $\le 10$.
\end{example}

We will now show the following lemma.

\begin{lemma}\label{lem:dodgsonscore}
$\phi$ is satisfiable if and only if DodgsonScore$(q) \le 4m +n$
\end{lemma}

\begin{proofs}
First suppose that $\phi(z_1,\ldots,z_n)$ is satisfiable. Fix an assignment $\alpha \in \{0,1\}^n$ for
$\phi$. We argue as in Example~\ref{ex:dod-3sat}.

Recall that $q$ needs one vote over each $\widehat{z}_i$, each $c_i$, and each $c_{i,j}$ candidate.
For each $t, 1 \le t \le n$ if $z_t$ is false in the assignment (i.e.,
$\alpha_t = 0$) then for each $c_{i,j}$ such that $\ell_{i,j} = z_t$,
swap $q$ with $c_{i,j}$ in the corresponding votes in Block~I
and swap $q$ up to the top in the corresponding $z_t$ vote in Block~II, and if
$z_t$ is true in the assignment (i.e., $\alpha_t = 1$) then for each $c_{i,j}$
such that $\ell_{i,j} = \overline{z_t}$, swap $q$ with $c_{i,j}$ in the
corresponding votes in Block~I and swap $q$ up to the top in the corresponding
$\overline{z_t}$ vote in Block~II.
For example, if $z_t$ is false in the satisfying assignment do the following.
\begin{itemize}
   \item Swap $q$ with $c_{i,j}$ in each Block~I vote 
     $(c_i > c_{i,j} > q > \cdots)$ for which $\ell_{i,j} = \overline{z_t}$
    to get $(c_i > q > c_{i,j} > \cdots)$.
   \item Swap $q$ up to the top in the Block~II vote
    $(\widehat{z}_t > \{c_{i,j}\ |\ \ell_{i,j} = z_t\} > q > \cdots)$ to get
   $(q > \widehat{z}_t > \{c_{i,j}\ |\ \ell_{i,j} = z_t\} > \cdots)$.
\end{itemize}
This takes $3m+n$ swaps.

Since $\alpha$ is a satisfying assignment for $\phi$ we know that for each clause there
exists a clause literal
that is true. In our election, this means that for each $i, 1 \le i \le m$,
there exists a $j, 1 \le j \le 3$, such that $q$ has been swapped in the vote
$(c_i > c_{i,j} > q > \cdots)$ to get $(c_i > q > c_{i,j} > \cdots)$.
So with a total of $m$ swaps
$q$ can gain one vote over each clause candidate $c_i$.
This all takes a total of $4m+n$ swaps.

For the converse, suppose that DodgsonScore$(q) \le 4m+n$.
The Dodgson score of $q$ will always be at least $4m+n$, since $q$
must gain one vote over $4m+n$ candidates (each $c_i$, each $c_{i,j}$, and each $\widehat{z}_i$ candidate).
To avoid ``wasting'' swaps, $q$ must swap only with these candidates
and only in Block~I and Block~II, since these are
the only votes were $q$ is directly adjacent to these candidates.
Since each swap must gain a necessary vote, for each $t, 1 \le t \le n$, we either swap $q$ up
in the Block~II vote $(\widehat{z}_t > \{c_{i,j}\ |\ \ell_{i,j} = z_t\} > q > \cdots)$ to get 
$(q > \widehat{z}_t > \{c_{i,j}\ |\ \ell_{i,j} = z_t\} > \cdots)$  or swap $q$ up in the Block~II vote
$(\widehat{z}_t > \{c_{i,j}\ |\ \ell_{i,j} = \overline{z_t}\} > q > \cdots)$ to get 
$(q > \widehat{z}_t > \{c_{i,j}\ |\ \ell_{i,j} = \overline{z_t}\} > \cdots)$.
Note that this is the only way for $q$ to gain one vote over
$\widehat{z}_t$ without wasting swaps.
In the first case set $z_t$ to false and in the second case set
$z_t$ to true. This gives us a satisfying assignment for $\phi$ since for each
$i, 1 \le i \le m$, $q$ must also swap with the clause candidate $c_i$, which implies that
for some $j, 1 \le j \le 3$, $q$ swaps with $c_{i,j}$ in Block~I, and since $q$ only swaps
over each of these candidates exactly once, $q$ did not swap with this $c_{i,j}$ in Block~II.
So $c_{i,j}$ corresponds to a true literal in the constructed satisfying assignment.~\end{proofs}

We will now use the idea from the reduction above to show that
Dodgson-\ccdcstar\ %
is \sigmatwo-complete.

Membership in \sigmatwo\ follows from Corollary~\ref{cor:upperbound}.
To show hardness, we reduce from QSAT$_2$ and recall that it is defined
as all true formulas of the form
$\exists x_1 \cdots \exists x_n \neg (\exists y_1 \cdots \exists y_n 
\phi(x_1, \ldots, x_n, y_1, \ldots, y_n))$, where $\phi$ is a formula in 3cnf,
where $\phi = \psi_1 \wedge \psi_2 \wedge \cdots \wedge \psi_m$ and
for each $i, 1 \le i \le m$, $\psi_i = \ell_{i,1} \vee \ell_{i,2} \vee \ell_{i,3}$.
Let $X = \{x_1, \ldots, x_n\}$ and $\overline{X} = \{\overline{x_1}, \ldots, \overline{x_n}\}$
denote the positive and negative $x$-literals and let
$Y = \{y_1, \ldots, y_n\}$ and $\overline{Y} = \{\overline{y_1}, \ldots, \overline{y_n}\}$
denote the positive and negative $y$-literals.
Without loss of generality, let $n > m$ and $n > 6$.
Let $\widehat{m}$ be the number of occurrences of $y$-literals, i.e., $\widehat{m} = \|\{(i,j)\ |\ 1 \le i \le m, 1 \le j \le 3,\ {\rm and}\ \ell_{i,j} \in Y \cup \overline{Y}\}\|$. Note that
$\widehat{m} \le 3m$.

We now construct an instance of Dodgson-\ccdcstar. Let
the candidate set $C$ consist of the $x$-variable
candidates $\{\widehat{x}_1, \dots, \widehat{x}_n\}$, the $y$-variable
candidates $\{\widehat{y}_1, \dots, \widehat{y}_n\}$,
the clause candidates $\{c_1, \dots, c_m\}$, a clause-literal candidate for
each occurrence of a $y$-literal $\{c_{i,j}\ |\ 1 \le i \le n, 1 \le j \le 3,\ {\rm and}\ \ell_{i,j} \in Y \cup \overline{Y}\}$ (note that there are $\widehat{m}$ clause-literal candidates),
the buffer candidates $\{b_1, b_2, b_3, b_4\}$,
the positive and negated $x$-literal candidates $X \cup \overline{X}$
and the candidates $p$, $q$, and $d$.
Let $p$ be the preferred candidate of the chair,
let $X \cup \overline{X}$ be the set of deletable candidates, and let the delete limit be $2n$ (i.e., we can delete any subset of $X \cup \overline{X}$;
this will be useful when we look at CCAC). 
We have the following $16n+8m+2\widehat{m}-8$ voters.

\begin{description}
    \item[Block IA] For each $i, 1 \le i \le m$, and  $j, 1 \le j \le 3$,
    such that $\ell_{i,j} = y_t\ {\rm or}\ \overline{y_t}$,
    \begin{itemize}
        \item One voter voting: $(c_i > c_{i,j} > q > b_1 > p > d > \cdots)$.
    \end{itemize}
	
\item[Block IB] For each $i, 1 \le i \le m$, and  $j, 1 \le j \le 3$,
such that $\ell_{i,j} = x_t\ {\rm or}\ \overline{x_t}$,
    \begin{itemize}
        \item {If $\ell_{i,j} = x_t$, one voter voting:} $(c_i > x_t > q > b_1 > p > d > \cdots)$.    	
        \item {If $\ell_{i,j} = \overline{x_t}$, one voter voting:} $(c_i > \overline{x_t} > q > b_1 > p > d > \cdots)$.
    \end{itemize}
\item[Block II] For each $t, 1 \le t \le n$,
    \begin{itemize}
      \item {One voter voting:} $(\widehat{y}_t > \{c_{i,j}\ |\ \ell_{i,j} = y_t\} > q > b_1 > p > d > \cdots)$.
      \item {One voter voting:} $(\widehat{y}_t > \{c_{i,j}\ |\ \ell_{i,j} = \overline{y_t}\} > q > b_1 > p > d > \cdots)$.
    \end{itemize}
\item[Block III] For each $t, 1 \le t \le n$,
    \begin{itemize}
      \item {One voter voting:} $(\widehat{x}_t > x_t > p > q > d > \cdots)$.
      \item {One voter voting:} $(\widehat{x}_t > \overline{x_t} > p > q > d > \cdots)$.
    \end{itemize}
\item[Block IV] \hfill
\begin{itemize}
      \item {One voter voting:} $(d > \cdots > b_2 > \{c_1, \ldots, c_m\} > \{{\rm all}\ c_{i,j}\} >\\ \{\widehat{y}_1, \ldots, \widehat{y}_n\} > p > b_1 > q).$
      \item One voter voting: $(d > \cdots > b_2 > \{\widehat{x}_1, \dots, \widehat{x}_n\} > q > b_1 > b_3 > p)$.
 \end{itemize}
\item[Block V]\hfill
    \begin{itemize}
        \item {One voter voting:} $(d > \cdots > b_2 > p > q > \{c_1,\ldots, c_m\})$.
        \item {$2n+3m-6$ voters voting:} $(d > \cdots > b_2 > p > b_1 > q)$.
    \end{itemize}
\item[Block VI]\hfill
    \begin{itemize}
        \item {$4n+m+\widehat{m}-2$ voters voting:} $(p > q > d > \cdots)$.
        \item {$2n+m+\widehat{m}-4$ voters voting:} $(\cdots > b_3 > q > b_4 > d > p > b_1 >\\ b_2 > X \cup \overline{X})$.
        \item {$4n-1$ voters voting:} $(d > \cdots > b_2 > q > b_1 > p > b_3 >b_4 > X \cup \overline{X})$.
         \item {Two voters voting:} $(\cdots > b_3 > d > p > b_4 > q > b_1 > b_2 > X \cup \overline{X})$.
    \end{itemize} 
\end{description}
Note the following about the election above.
\begin{enumerate}
	\item $q$ needs one vote over each $\widehat{x_i}$, each $\widehat{y}_i$, each $c_i$, 
    each $c_{i,j}$, and $p$, and no votes over other candidates.
    Note that DodgsonScore$(q) \ge 2n+m+\widehat{m}+1$.
\item %
	The votes in Block~IA %
	and the Block~II votes function in the
same way to determine (part of) the score of $q$ as in the Dodgson score reduction above.
And $q$ can only gain a vote over any of the
 $\widehat{y_i}$, $c_i$, or $c_{i,j}$
candidates without wasting a swap in Block~I and Block~II.
	\item \label{dod:con:p} $p$ needs one vote over each $\widehat{x}_i$, each $\widehat{y}_i$,
    each $c_i$, each $c_{i,j}$, and $q$, and no votes over other candidates.
    $p$ needs at least two swaps to gain one vote over $q$ and to accomplish this $p$
    needs to swap over a buffer candidate.
    Note that DodgsonScore$(p) \ge 2n+m+\widehat{m}+2$.

\item $d$ needs $2n+m+\widehat{m}-1$ votes over $q$, 3 votes over $p$, and no votes
    over other candidates. Note that regardless of
which candidates in $X\cup \overline{X}$ are deleted, DodgsonScore$(d) = 2n+m+\widehat{m}+2$.

    \item Each candidate in $C-\{p,q,d\}$ needs at least $2n+m+\widehat{m}+2$ votes over $d$. So
    regardless of which candidates in $X \cup \overline{X}$ are deleted, for each $c \in C-\{p,q,d\}$, DodgsonScore$(c) \ge 2n+m+\widehat{m}+2$.

\end{enumerate}

Let's consider the Dodgson score of $p$. %
It is easy to see for every
$X' \subseteq X \cup \overline{X}$, in the election $(C-X',V)$, $p$ can gain each needed
vote over the $\widehat{y_i}$, $c_i$, and $c_{i,j}$ candidates in the Block~IV vote
and then use two more swaps to gain one vote over $q$. %
What remains is to show
how $p$ can gain one vote over each of the $\widehat{x}_i$ candidates.
When at least one of the $\{x_i,\overline{x_i}\}$ candidates is deleted,
$p$ can gain one vote over the corresponding $\widehat{x}_i$ candidate
in Block~III, using one swap.
So, we can state the following
observation.

\begin{observation}\label{obs:ccdcstar0}
For every $X' \subseteq X \cup \overline{X}$, 
\begin{enumerate}
    \item The Dodgson score of $p$ with $X'$ removed is $\ge 2n+m+\widehat{m}+2$.
    \item The Dodgson score of $p$ with $X'$ removed is $2n+m+\widehat{m}+2$ if and
        only if for each $i, 1 \le i \le n$, at least one of each $x_i$ or $\overline{x_i}$ in is $X'$.
\end{enumerate}
\end{observation}

We now consider the Dodgson score of $q$.

\begin{lemma}\label{lem:ccdcstar1}
    For every $X' \subseteq X\cup \overline{X}$,
    \begin{enumerate}
        \item The Dodgson score of $q$ with $X'$ removed is $\ge 2n+m+\widehat{m}+1$.
        \item The Dodgson score of $q$ with $X'$ removed is $2n+m+\widehat{m}+1$ if and only if
            $\phi$ with $x$-literals in $X'$ set to true and $x$-literals not in $X'$ set
            to false is satisfiable.
    \end{enumerate}
\end{lemma}
\begin{proofs}
\begin{enumerate}
    \item Clearly the Dodgson score of $q$ with $X'$ removed is $\ge 2n+m+\widehat{m}+1$, since $q$ needs one vote over each $\widehat{x}_i$, each $\widehat{y}_i$, each $c_i$, each $c_{i,j}$, and $p$,
and does not need any votes over other candidates.

\item Suppose that $q$ can be made a Condorcet
winner in the election $(C-X',V)$
with exactly $2n+m+\widehat{m}+1$ swaps.
This means that every swap must gain a necessary vote, since $q$ needs one
vote each over $2n+m+\widehat{m}+1$ candidates. With the exception of candidate $p$ and
the $\widehat{x}_i$ candidates,
$q$ is only ever not separated from candidates it needs votes over by a
buffer candidate
in Block~I and Block~II.
So $n+m+\widehat{m}$ of these swaps must occur there. Note that $q$ can easily swap over $p$
(in Block III), and the $\widehat{x}_i$ candidates (in Block~IV) with exactly $n+1$ swaps.

Consider the candidates deleted in $X'$. If $\ell_{i,j} = x_t\ (\overline{x_t})$ then
$q$ can gain one vote over $c_i$ in the corresponding Block~I vote with only one swap.
This will correspond
to setting $x_t$ to true (false) in the corresponding assignment and
setting $x_t\ (\overline{x_t})$ to true means that the $i$th clause is true.

Now let's consider the $y$-literals.
Since the Dodgson score of $q$ is $2n+m+\widehat{m}+1$, $q$ can swap over the
$\widehat{y}_i$, $c_i$, and $c_{i,j}$ candidates 
without wasting swaps. It follows from
a similar argument as in the proof of Lemma~\ref{lem:dodgsonscore} that since $q$ can
swap over the remaining $n+m+\widehat{m}$ candidates
without wasting a swap, $\phi$ with the $x$-literals in $X'$ set to true
and $x$-literals not in $X'$ set to false is satisfiable.
For the other direction,
consider a fixed satisfying assignment for
$y$-variables in $\phi$, where the assignment to the $x$-literals is
set by the deleted candidates $X'$.
For each $t$, $1 \le t \le n$, if $y_t$
is true (false) in the satisfying assignment,
swap $q$ up over the $c_{i,j}$ candidates and
$\widehat{y}_t$ in the $\overline{y_t}\ (y_t)$ vote in Block~II. Then for each $t$, $1 \le t \le n$,
if $y_t$ is true (false) swap $q$ with $c_{i,j}$ in the corresponding $\ell_{i,j} = y_t\ (\overline{y_t})$
Block~I votes.
This takes $n+m$ swaps and handles
the $c_{i,j}$ and $\widehat{y}_i$ candidates. Since we are looking at a satisfying assignment,
for all $i$, there exists a $j$, $1 \le j \le 3$ such that $c_i > c_{i,j} > q$ has been
swapped to $c_i > q > c_{i,j}$ or $x_t\ (\overline{x_t})$ has been deleted so that $c_i > q$.
With $k$ extra swaps, $p$ beats $c_i$ pairwise
for all $i$. $q$ must additionally gain one vote over
$p$ and one vote over each $\widehat{x}_i$ candidate and it is clear to see how this can be
accomplished with $n+1$ swaps.
This all takes $2n+m+\widehat{m}+1$ swaps.
\end{enumerate}\end{proofs}

We will now show that $\exists x_1 \cdots \exists x_n \neg (\exists y_1 \cdots \exists y_n 
\phi(x_1, \ldots, x_n, y_1, \ldots, y_n)) \in \qsattwo$ if and only
if there exists $X' \subseteq X \cup \overline{X}$
such that $p$ is a Dodgson winner of $(C-X', V)$.

Suppose that $\exists x_1 \cdots \exists x_n \neg (\exists y_1 \cdots \exists y_n \phi(x_1, \ldots, x_n,\\ y_1, \ldots,
y_n) \in \qsattwo$.
Fix an assignment $\alpha \in \{0,1\}^n$ for the $x$-literals such that $\phi$ is not satisfiable. Let the 
candidates deleted be
$X' = \{x_i\ |\ \alpha_i =\ 1\} \cup \{\overline{x_i}\ |\ \alpha_i = 0\}$.
We know from Observation~\ref{obs:ccdcstar0} that the Dodgson score of $p$ with $X'$ deleted
is $2n+m+\widehat{m}+2$ and by Lemma~\ref{lem:ccdcstar1} we know that DodgsonScore$(q) > 2n+m+\widehat{m}+1$.
 And as stated in the list of properties given after the construction, the Dodgson score of
each candidate in $C-\{p,q,d\}$ is at least $2n+m+\widehat{m}+2$.
Additionally, the Dodgson score of $d$ is $2n+m+\widehat{m}+2$ regardless of which candidates in
$X \cup \overline{X}$ are deleted. It follows that $p$ is a Dodgson winner of $(C-X',V)$.

Suppose that there exists an $X' \subseteq X \cup \overline{X}$
such that 
$p$ is a winner. Regardless of which candidates in $X \cup \overline{X}$ are deleted,
DodgsonScore$(d)=2n+m+\widehat{m}+2$ and DodgsonScore$(p) \ge 2n+m+\widehat{m}+2$.
Since $p$ is a winner, DodgsonScore$(p) = 2n+m+\widehat{m}+2$ and it follows from
Observation~\ref{obs:ccdcstar0} that $X'$ contains at least
one of each $\{x_i,\overline{x_i}\}$.
Let $\alpha \in \{0,1\}^n$ be an assignment to the $x$-literals that
is consistent with the candidates
in $X'$, that is, for all $i, 1 \le i \le n$, if
$\alpha_i = 1$, then $x_i \in X'$
and if $\alpha_i = 0$, then
$\overline{x_i} \in X'$.
Suppose that $\phi(\alpha_1, \ldots, \alpha_n, y_1, \ldots, y_n)$ is satisfiable.
Then by Lemma~\ref{lem:ccdcstar1}, DodgsonScore$(q) = 2n+m+\widehat{m}+1$.
However, since DodgsonScore$(p) = 2n+m+\widehat{m}+2$, $p$ is not a winner, which is a contradiction.
It follows that $\phi(\alpha_1, \ldots, \alpha_n, y_1, \ldots, y_n)$ is not satisfiable.~\end{proofs}

The analogous result for CCAC follows almost immediately.

\begin{theorem}
Dodgson-CCAC is \sigmatwo-complete.
\end{theorem}

\begin{proofs}
Membership in \sigmatwo\ follows from Observation~\ref{obs:control}.
To show hardness, we reduce from QSAT$_2$ and use the same construction as for
the case of the proof of Theorem~\ref{dod:ccdcstar}, with the obvious modification
that the set of deletable candidates ($X \cup \overline{X})$ is now the set of
unregistered candidates and the delete limit
is now the addition limit of $2n$. Note that $p$ can be a winner by 
deleting a set of candidates $\widehat{X} \subseteq X \cup X'$ 
from $(C,V)$ if and only if
$p$ can be made a winner by adding $(X \cup X') - \widehat{X}$
to $(C - (X \cup X'), V)$,  which immediately gives the result.
\end{proofs}

The construction used for the proof of Dodgson-\ccdcstar\ can be 
adapted to show the following.

\begin{theorem}
Dodgson-CCDC is \sigmatwo-complete.
\end{theorem}

\begin{proofsketch}
Membership in \sigmatwo\ follows from Corollary~\ref{cor:upperbound}.
We will adapt the construction used in the
proof of Theorem~\ref{dod:ccdcstar}.  Set the delete limit to $n$.
Our main challenge is to ensure that the arguments from the previous
proof still go through when we delete up to $n$ arbitrary candidates. 
We start with Blocks~I,~II, III, and~IV and do the following.
\begin{enumerate}
    \item To ensure that the buffer candidates function in the same way, we
    replace each buffer candidate by $n+1$ copies of that candidate.

    \item We also replace the candidate ``$d$,'' which in the Dodgson-\ccdcstar\ proof, regardless of which candidates in $X \cup \overline{X}$ are deleted,
always has Dodgson score equal to the minimum-possible Dodgson score of $p$,
    by $n+1$ copies:
    $\{d_1, \dots, d_{n+1}\}$.
    And we replace each occurrence of $d$ in the construction
    by $d_1 > d_2 > \cdots > d_{n+1}$. This  will ensure that regardless of which candidates
    in $C-\{q\}$ are deleted, the Dodgson score of the first $d_i$ that is not
    deleted is the minimum-possible Dodgson score of $p$. (We will mention why
    we do not delete $q$ later.)

    \item We now will make sure that the candidates deleted correspond to an
    assignment. In the Dodgson-\ccdcstar\ proof, this was accomplished by
    the Block~III votes and the $\widehat{x}_i$ candidates.
\begin{description}
    \item[Old Block III] For each $t, 1 \le t \le n$,
    \begin{itemize}
      \item \indent {One voter voting:} $(\widehat{x}_t > x_t > p > \cdots)$.
      \item {One voter voting:} $(\widehat{x}_t > \overline{x_t} > p > \cdots)$.
    \end{itemize}
\end{description}
In our new construction we create $2n$ of each of the $\widehat{x}_i$ candidates:\\
$\{\widehat{x}_1^1,\dots,\widehat{x}_1^{2n}, \dots, \widehat{x}_n^{1}, \dots, \widehat{x}_n^{2n}\}$ and we make sure that $p$ needs one vote over
each of these $2n^2$ variables.
We then change Block~III to be the following $4n^2$ votes.
\begin{description}
\item[New Block III] For each $t, 1 \le t \le n$, and $s, 1 \le s \le 2n$,
    \begin{itemize}
      \item {One voter voting:} $(\widehat{x}_t^s > x_t > p > \cdots)$.
      \item {One voter voting:} $(\widehat{x}_t^s > \overline{x_t} > p > \cdots)$.
    \end{itemize}
\end{description}
Now for each $x_i$ or $\overline{x_i}$ deleted, $p$ can gain
the necessary vote over each of the at least $n$
nondeleted $\widehat{x}_i^s$ candidates without wasting swaps, and
so one such deletion decreases the Dodgson score of $p$ by at least $n$.
Note that
if for an $i$, $1 \le i \le n$, both of $x_i$ and $\overline{x_i}$ were deleted, or if
a candidate in $C - (X \cup \overline{X})$ was deleted, then the Dodgson score of $p$
would decrease by at most 1.

In order for $p$ to tie with ``$d$'' by deleting at most $n$ candidates, we need to
delete exactly one of each $\{x_i,\overline{x_i}\}$, which corresponds to an
assignment. Note that this also implies that we do not delete $q$.

\item %
In the Dodgson-\ccdcstar\ proof, to get its needed vote over $q$,
$p$ needs to waste a swap over $q$ and does this by swapping over a buffer candidate.
Since there are now $n+1$ copies of each buffer candidate, we need to modify the votes
so that $p$ still wastes exactly one swap to gain its vote over $q$.
Recall the Block~II votes from the Dodgson-\ccdcstar\ proof.
\begin{description}
\item[Old Block II] For each $t, 1 \le t \le n$,
    \begin{itemize}
        \item {One voter voting:} $(\widehat{y}_t > \{c_{i,j}\ |\ \ell_{i,j} = y_t\} > q > b_1 > p > d > \cdots)$.
        \item {One voter voting:} $(\widehat{y}_t > \{c_{i,j}\ |\ \ell_{i,j} = \overline{y_t}\} > q > b_1 > p > d > \cdots)$.
    \end{itemize}
\end{description}
In our new construction, we replace the buffer candidate with candidates from
$X \cup \overline{X}$ in the following way.
\begin{description}
\item[New Block II] For each $t, 1 \le t \le n$,
    \begin{itemize}
        \item {One voter voting:} $(\widehat{y}_t > \{c_{i,j}\ |\ \ell_{i,j} = y_t\} > q > \{x_t,\overline{x_t}\} > p > d > \cdots)$.
        \item {One voter voting:} $(\widehat{y}_t > \{c_{i,j}\ |\ \ell_{i,j} = \overline{y_t}\} > q > \{x_t,\overline{x_t}\} > p > d > \cdots)$.
    \end{itemize}
\end{description}
Since from above we know that an assignment must be deleted, we know that one of
each $\{x_i,\overline{x_i}\}$ remains after deletion, and so to gain one vote
over $q$, $p$ swaps over a remaining $x$-literal candidate in a New Block~II vote
to then swap over $q$. Note that in votes outside of Block~II where $p$ is ranked
below $q$, $p$ is separated from $q$ by at least $n+1$ buffer candidates.

\item We must then pad the above construction
(Blocks I-IV) so that the following hold.
    \begin{itemize}
        \item $q$ needs one vote
    over each $\widehat{x}_i^j$, each $\widehat{y}_i$, each $c_i$, 
    each $c_{i,j}$, and $p$, and no votes over other candidates.
    \item $q$ can only gain a vote over any of the $\widehat{y_i}$, $c_i$, or $c_{i,j}$ candidates without wasting a swap in
Block~I and Block~II.
	\item $p$ needs one vote over each $\widehat{x}_i^j$, each $\widehat{y}_i$,
    each $c_i$, each $c_{i,j}$, and $q$, and no votes over other candidates.
$p$ wastes a swap to gain a vote over $q$.
    \item Each $d_i \in D$ needs $2n^2+n+m+\widehat{m}+2$ votes in total over
    $p$ and $q$, %
    and no votes over candidates in $C-(\{p,q\} \cup D)$.
$d_1$ can gain these $2n^2+n+m+\widehat{m}+2$ votes without wasting swaps.
    \item Each candidate in $C-(\{p,q\}\cup D)$ needs at least $2n^2+n+m+\widehat{m}+2$
    votes over the candidates in $D$.
    \end{itemize}
\end{enumerate}
\end{proofsketch}

\end{document}